\documentclass[aps,physrev,superscriptaddress]{revtex4-2}

\usepackage{graphicx}
\usepackage{dcolumn}
\usepackage{bm}
\usepackage{color}
\usepackage[normalem]{ulem}
\usepackage{soul}
\usepackage{mathtools}

\def\CITE#1{{\color{red}[CITE]}}
\newcommand{\KK}{\mathcal{K}}
\newcommand{\DD}{D_{\mathrm{eff}}}

\usepackage{hyperref}

\begin{document}


\title{Phase dynamics and their role determining energy flux in hydrodynamic shell models}

\author{Santiago J. Benavides}
\email[]{Santiago.Benavides@ed.ac.uk}
\affiliation{School of Aeronautics and Space Engineering, Universidad Polit\'{e}cnica de Madrid, Madrid, Spain}
\affiliation{Current address: School of Mathematics and Maxwell Institute for Mathematical Sciences, University of Edinburgh, Edinburgh, UK}

\author{Miguel D. Bustamante}
\email[]{miguel.bustamante@ucd.ie}
\affiliation{School of Mathematics and Statistics, University College Dublin, Belfield, Dublin 4, Ireland}

\date{\today}

\begin{abstract}
The transfer of energy and other conserved quantities across scales, also known as flux or spectral flux, is a central aspect of out-of-equilibrium systems such as turbulent hydrodynamic flows. Despite its role in the few predictive theories that exist, a dynamical understanding of what determines said flux (and its direction in scale) has yet to be established. In this study, we work towards this understanding by investigating how the dynamics of complex Fourier velocity phases influence the flux of conserved quantities in hydrodynamic shell models. The phase dynamics, like energy evolution, are influenced by contributions from all neighboring triads, making the full problem intractable. Instead, we assume that the dynamics of the triad phases are determined solely by the so-called self-interaction term and treat the other neighboring triad terms as noise. This transforms the phase dynamics into that of a noisy phase oscillator, which we solve analytically to predict phase statistics. We validate this assumption with a suite of shell model simulations. Our results give us analytical predictions for the energy flux, when the energy spectrum is given. We prove that all shell models that conserve energy along with a sign-indefinite quadratic quantity (this includes three-dimensional turbulence analogues) undergo a forward energy cascade, and further show that the phase dynamics prevent the shell model analogue of two-dimensional turbulence from forming an inverse cascade of energy.
\end{abstract}


\maketitle

\section{Introduction \label{sec:intro}}
\subsection{Motivation \label{subsec:motivation}}
Turbulent flows are characterized by a large scale-separation between energy injection and dissipation. The energy injected is transferred among scales, via a so-called inertial range, until it reaches dissipation scales. How this energy transfer develops can determine the emergent properties of the flow, such as the size and lifetimes of different flow structures formed \cite{FrischBook,JimenezReview,DavidsonBook,AlexakisReview}. 
In three-dimensional (3D) hydrodynamic flows the energy injected cascades from large spatial scales to smaller scales, in what is known as a forward energy cascade \cite{FrischBook,VermaBook}. In two-dimensional (2D) hydrodynamic flows the cascade direction reverses and the energy undergoes a so-called inverse energy cascade, namely from small spatial scales to larger scales \cite{BoffettaReview,AlexakisReview}. In geophysical and astrophysical settings, where the turbulence is three-dimensional but anisotropic, the energy can simultaneously cascade to both large and small scales depending on the relevant geophysical parameter, such as rotation, stratification, aspect ratio, or magnetic field strength \cite{DavidsonBook,AlexakisReview,VermaBook,MarstonReview}. 
The direction of the energy cascade can have significant consequences on properties of interest, such as the mixing efficiency of passive scalars in geophysical settings,
error propagation in weather prediction,
and climate model closures \cite{Mellor1982,KalnayBook,Wunsch2004,Dauxois2021,Storer2022}.
Despite the fundamental importance of the energy cascade in turbulent flows, 
a predictive theory of the energy cascade direction in turbulent flows, and its model analogues, is still lacking.

Much work has been done to uncover the physical mechanisms responsible for inter-scale energy transfer, resulting in the identification of vortex stretching and strain self-amplification in 3D \cite{Eyink2006MSG,Doan2018,Carbone2020,Johnson2020,Johnson2021,JohnsonReview,Park2025}
and vortex clustering and vortex thinning by large-scale strain in 2D \cite{Kraichnan1967,Kraichnan1971,Mcwilliams1990,Chen2006,Eyink20062DCL,Xiao2009,Jimenez2021}
as primary drivers of the energy cascade. These findings are largely a result of analyses on coarse-grained direct numerical simulations (DNS) in physical space, with a focus on the roles (and alignment \cite{Chen2006,Ballouz2018,Ballouz2020}) of the sub-grid stress and large-scale strain tensors. 
While this approach does provide plausible explanations for the observed direction of the energy cascade, based on ``benchmark'' mechanisms or flow regimes such as the alignment of vortex lines with the second eigendirection of the strain tensor (supported by DNS and by simplified toy models such as Vieillefosse's \cite{Vieillefosse1982,Vieillefosse1984}), it does not yet provide a dynamical mechanism of general use to predict cascade directions. It therefore excludes any dynamical considerations which might explain, for example, the cascade reversals observed in anisotropic turbulent flows \cite{AlexakisReview,vanKan2024}. 
 
Other approaches focus on the Fourier space (`spectral') representation of the Euler or Navier Stokes equations.
One of the earliest such studies by Fj{\o}rtoft considers the dissipation scale of positive-definite conserved quantities, and can be generalized to other models of turbulent flows \cite{Fjortoft1953,AlexakisReview}. It can be used to show that no forward energy cascade can coexist with a forward cascade of enstrophy, the volume-averaged squared vorticity, in 2D turbulence. However, this argument can only be used to exclude possible cascade directions, but does not dynamically predict whether a cascade exists or not. Indeed, in 3D turbulence there is only one positive-definite conserved quantity, and the energy cascade direction is unconstrained. Fj{\o}rtoft's arguments also assume the presence of an energy cascade, which, as we will discuss in this paper, is not always a given for some models of turbulence. 

At around the same time, statistical physics approaches were applied to the truncated Euler equations, resulting in predictions for absolute statistical equilibrium states of 2D and 3D turbulence \cite{Hopf1952,Lee1952,Kraichnan1973,Kraichnan1967,Kraichnan1980,AlexakisReview}. These are states which satisfy detailed balance and have no energy flux, differing greatly from the out-of-equilibrium turbulent states found when viscosity is nonzero. Despite this, it has been argued that an out-of-equilibrium system seeks its equilibrium state, and that this determines the energy cascade direction \cite{Kraichnan1967,Kraichnan1973,Kraichnan1975,Kraichnan1980}. Unlike the Fj{\o}rtoft argument, it predicts the forward cascade of energy in 3D isotropic turbulence, since the absolute equilibrium state for 3D truncated Euler has more energy at smaller scales than does its out-of-equilibrium counterpart. A downside to this approach, however, is that it still depends on the system's conserved quantities. This becomes a problem for anisotropic flows in geophysical settings, such as rotating 3D turbulence, which has the same conserved quantities as the non-rotating case. Unless one restricts the analysis to a particular subspace of phase space \cite{Bourouiba2008,Herbert2014}, equilibrium theory predicts a forward cascade, despite the clear observation of an inverse cascade at large rotation rates \cite{Smith1999,vanBokhoven2009,Yarom2013,Buzzicotti2018}.

Another approach worth noting is that of the `instability assumption' of Waleffe \cite{Waleffe1992,Waleffe1993}, who considered the inviscid dynamics of an isolated triad of wave vectors. He conjectured that the stability properties of an isolated triad reflect the triple correlations in a fully turbulent flow, responsible for energy transfers among scales. For example, homochiral triads (each mode's helicity being of the same sign) have an unstable middle wavenumber, resulting in energy flowing to larger and smaller wavenumbers. The instability assumption then states that these triads are responsible for an inverse cascade of energy. Indeed, homochiral triads in 3D turbulence have been found to produce an inverse energy flux \cite{Biferale2012,Biferale2013,Alexakis2017,Sahoo2017,Alexakis2022},
and all triads in 2D turbulence have the same stability properties \cite{Waleffe1992}. The dynamical basis of this approach opens the door to analysis of anisotropic systems, such as rotating flows \cite{Waleffe1993} where the instability assumption correctly predicts a transfer of energy towards (but not directly to) the modes that do not vary along the rotation axis, as observed in DNS \cite{Buzzicotti2018}. 

Why the instability assumption produces correct predictions is still an open question, particularly when considering the fact that the dynamics of a single triad represent a 2D three-component system which has more conserved quantities and different dynamics than that of 3D turbulence \cite{Moffatt2014,Biferale2017}. Furthermore, none of the approaches described above predict how the fraction of energy going to large scales depends on the geophysically-relevant control parameter in anisotropic flows, let alone  capture the apparent phase transition in the cascade which is observed \cite{Seshasayanan2014,Seshasayanan2016,Benavides2017,Pestana2019,AlexakisReview,VanKan2020,vanKan2024}.
Our goal in this study is to take the next step towards an understanding of the energy cascade in turbulence (and its models) by introducing a new approach based on the dynamics of the complex phases of the Fourier amplitudes in spectral space. As a first step, we will uncover an approximate statistical description of these phases in shell models of hydrodynamic turbulence, which we will use to make new predictions about the cascade of conserved quantities.

\subsection{Background \label{subsec:background}}
In the spectral representation of the Euler or Navier Stokes equations, one works with the complex-valued amplitudes of the velocity field, which are evaluated for each wave vector $\vec{k}$ or, more informally, each mode. Despite most of the focus in turbulence on the magnitudes of these amplitudes, which relate to the energy spectrum and length and time scales of the flow, the complex phase of the velocity amplitudes play a crucial role in the dynamics, and in particular the transfer of energy and other conserved quantities across scales. 
This can be seen from the evolution of the energy of a mode $\vec{k}$, from which an energy flux can be computed. Although the complex phase of the mode in question cancels out in all linear terms due to the multiplication by its complex conjugate, the nonlinear term becomes a triple product of complex amplitudes for all the triads that can form with $\vec{k}$. The complex phases, which sum together to form what is called a `triad phase', do not cancel out, and their statistical behavior determine the magnitude and direction of the energy flux, as we will write out explicitly in Section \ref{sec:phase_dyn_shell}.

The statistics of the complex phases has been the focus of recent studies of the one-dimensional (1D) Burgers equation, which  show that forward energy flux is associated with alignments of the triad phases, so that the most probable value of a triad phase becomes $\pi/2$, and with synchronization of the triad phases, so that triad phases across scales are positively correlated \cite{Buzzicotti2016,Moradi2017,Murray2018,Arguedas2022,Protas2024,Wang2024}. 
Although more complicated, a few studies have looked at triad phases in 2D and 3D flows, and have shown their relevance in extreme energy transfer events \cite{Kang2021}, dissipative structures \cite{Reynolds2016}, and decay \cite{Wang2024}.
Indeed, it can be easily seen from the equations of motion that a random-uniform triad phase (implying uncorrelated individual phases) results in no net energy transfer on average. However, a description of the statistics of the triad phase has yet to be formulated, due to the fact that their evolution equation contains interactions with all neighboring triad terms, making the dynamics complex and highly nonlinear. 

For this reason, we first focus on studying a reduced model of turbulence, called a shell model \cite{Biferale2003Review,DitlevsenBook}. These are models of turbulence which make three major simplifications to the spectral representation, thereby making analysis more approachable and a direct computation of their dynamics more efficient: (i) spectral space is divided into \textit{shells} enumerated by a single index $n$ and spaced apart by a factor $\lambda (>1)$, typically set to $2$. This means that the modes of the system are $k_n \coloneq k_0 \lambda^n$. (ii) There is a single complex velocity amplitude for each mode, that is $u_n \coloneq u(k_n)$. (iii) only \textit{local} interactions are considered in the nonlinear term, so that, what would normally be sum over all triads, becomes a sum over nearby neighbors. The equations and formulation of the specific shell model in this study will be introduced in Section \ref{subsec:HD_shell}. Although 1D Burgers is also a one-dimensional system, its nonlinear term still considers all possible triads of the system and hence is highly nonlocal in its formulation. Another important advantage of shell models is that one can choose the conserved quantities of the system, facilitating the study of systems with both forward and inverse cascades and going beyond the behavior of 1D Burgers.

Shell models of 3D turbulence, conserving energy and a helicity-like sign-indefinite second quantity, successfully reproduce the forward energy cascade as well as small-scale intermittency \cite{Biferale2003Review}.
Shell models for 2D turbulence, whose second invariant is an enstrophy-like positive-definite quantity, have not found as much success in reproducing turbulent statistics \cite{Aurell1994,Ditlevsen1996,Gilbert2002,Gurcan2016}.
Instead of finding the expected inverse cascade, a quasi-equilibrium state with almost zero energy flux was observed, leading to a primary focus on shell models of 3D turbulence in future studies. Apart from arguments based on equilibrium states \cite{Gilbert2002}, a dynamical understanding of the breakdown of the 2D turbulence model is lacking. 

The shell models' ability to provide quick and efficient computations of a turbulent-like system with good scale separation is ideal for intermittency studies, which require large quantities of time-series data taken from a well-defined inertial range.
This has led to them being extensively used to investigate the behavior and models of intermittency and its associated anomalous exponents \cite{Biferale2003Review,Mailybaev2021,Mailybaev2022,deWit2024},
but also position them as ideal for the study of energy transfers and cascades of conserved quantities across scales. Inspired by recent work on 1D Burgers \cite{Buzzicotti2016,Moradi2017,Murray2018,Arguedas2022,Protas2024}, analysis of triad phase statistics and their synchronization in shell models has revealed the importance of both triad phase alignment and synchronization in the transfer of energy \cite{CarrollThesis,Manfredini2025}. In fact, the significance of the triad phase in the inter-scale transfer of energy was already noted in shell model studies of intermittency \cite{Benzi1993}, long before its appearance in the study of 1D Burgers. 

In what follows, we make a crucial simplifying assumption on the dynamics of the triad phase for hydrodynamic shell models, resulting in an analytical approximation for the statistics of the triad phases. Given the new statistical description, we write, for the first time, an explicit expression for the average phase alignment and flux of conserved quantities for shell models as a function of the exponent of the kinetic energy spectrum as well as the second conserved quantity. These results are confirmed by numerical simulations of so-called phase-only shell models, where we fix the kinetic energy spectrum and evolve only the complex phases. Implications on the evolution of non-phase-only (i.e. `full') shell models, past observations of shell models, as well as on an extension to the full Navier-Stokes equations are discussed in the conclusions and in Appendix \ref{app:2D}.

\section{Phase dynamics of hydrodynamic shell models \label{sec:phase_dyn_shell}}

\subsection{Hydrodynamic shell models \label{subsec:HD_shell}}
We consider the dynamics of the `improved' shell model of L'vov \textit{et al.} \cite{Lvov1998}, defined to be:
\begin{eqnarray}
    \frac{d u_n}{dt} = &&i\left(a k_{n+1} u^*_{n+1} u_{n+2} + b k_n u^*_{n-1}u_{n+1}\right. \nonumber \\
    &&\left.- c k_{n-1} u_{n-2}u_{n-1}  \right) + f_n + D_n u_n,
\label{eq:full_eq}
\end{eqnarray}
where $k_n = k_0 \lambda^n$ is the wavenumber, $u_n(t) = u(k_n,t)$ is the complex velocity amplitude for each mode (with $u^*_n$ its complex conjugate), $f_n$ is a forcing, $D_n$ is a dissipative mechanism, and $[a,b,c]$ are coefficients which determine the conserved quantities of the system. The index $n$ goes from $0$ to $N-1$, where $N$ is the total number of shells in the system. The boundary conditions are $u_{-2} = u_{-1}=u_N=u_{N+1}=0$. The forcing and dissipative mechanism will remain unspecified for now.

Following standard practice, we choose $a = 1$. Energy conservation requires that $a+b+c = 0$, which we use to define $c = -a - b$. Our system then conserves the total energy, 
\begin{equation}
    E = \frac{1}{2} \sum_n |u_n|^2.
\end{equation}

This leaves only one free parameter, $b$, which determines the second conserved quantity, which we will call $H$. For $-1<b<0$, $H$ is sign-indefinite, making it analogous to 3D turbulent flows, and is of the form,
\begin{equation}
    H = \frac{1}{2} \sum_n  (-1)^n k_n^{\gamma_{3D}(b)} |u_n|^2,
\end{equation}
where 
\begin{equation}
    \gamma_{3D}(b) = \log_\lambda \left( \frac{1}{1+b}\right),
\end{equation} 
or, equivalently, $b(\gamma_{3D}) = \lambda^{-\gamma_{3D}}-1$, with $\gamma_{3D}>0$. The value of $\gamma_{3D} = 1$ corresponds to the analogy of Helicity in 3D flows. The shell models with an $H$ which is sign-indefinite will be called `3D-like', regardless of whether $\gamma_{3D} = 1$ or not.

On the other hand, for $-2 < b < -1$ the second conserved quantity is positive definite, making it analogous to 2D turbulent flows and is of the form,
\begin{equation}
    H = \frac{1}{2} \sum_n k_n^{\gamma_{2D}(b)} |u_n|^2,
\end{equation}
where 
\begin{equation}
    \gamma_{2D}(b) = \log_\lambda \left( \frac{-1}{1+b}\right), \label{eq:gamma_def}
\end{equation} 
or, equivalently, $b(\gamma_{2D}) = -\lambda^{-\gamma_{2D}} - 1$, with $\gamma_{2D}>0$. For $\gamma_{2D} = 2$, $H$ corresponds to an analogue of Enstrophy in 2D turbulence. As above, the shell models with an $H$ which is positive-definite  will be called `2D-like', regardless of whether $\gamma_{2D} = 2$ or not. 

For simplicity of exposition, for the remainder of the text we combine the two cases by defining,
\begin{equation}
    b(\gamma) = \pm \lambda^{-\gamma}-1, \label{eq:Ab_def}
\end{equation}
resulting in $c = \mp \lambda^{-\gamma}$ since $a = 1$.
Choosing the top sign results in 3D-like, sign-indefinite $H$ with exponent $\gamma$, whereas choosing the bottom sign results in 2D-like, positive-definite $H$ with exponent $\gamma$. Using this shorthand will simplify calculations in subsequent sections, particularly once we introduce the phase-only model in Section \ref{subsec:phase_only}.

Next we perform a change of variables, choosing a polar representation of the complex variable $u_n$:
\begin{equation}
    u_n \eqcolon \rho_n \mathrm{e}^{i \phi_n}, \label{eq:polar}
\end{equation}
where $\rho_n$ is the magnitude and $\phi_n$ is the phase of $u_n$, and similarly for the forcing, $f_n = \rho_n^f \mathrm{e}^{i \phi_n^f}$.
The evolution equations for each variable can be found by plugging in Eq. (\ref{eq:polar}) into Eq. (\ref{eq:full_eq}) and manipulating the expression so as to isolate $d \rho_n /dt$ and $d \phi_n /dt$. Doing so reveals that the evolution equations only depend on the \textit{triad phase},
\begin{equation}
    \theta_n \coloneq \phi_{n+2} - \phi_{n+1} - \phi_n,
\end{equation} 
and not on the individual phases (with the exception of the forcing term).
In fact, one can prove that the \emph{individual} phases $\phi_n$ are uniformly distributed \cite{Benzi1993}, yet an out-of-equilibrium solution requires that the \emph{triad} phases not be uniformly distributed, making it the dynamical variable of interest. We therefore perform one last change of variables, making the primary variables $(\rho_n,\theta_n)$. Note that, although the triad phase $\theta_n$ has a single index $n$, it comprises phases from three different modes: $n,n+1,$ and $n+2$.The single index is due to the fact that in shell models, by construction, any triad consists of three ``contiguous'' neighbors, and so one can choose for example the lowest shell index in the triad to ``represent'' the triad.

The evolution of $\rho_n$ becomes
\begin{eqnarray} 
    \frac{d \rho_n }{ d t } &=& - a k_{n+1} \rho_{n+2} \rho_{n+1} \sin\left(\theta_n\right) \nonumber \\ 
    && - b k_{n} \rho_{n+1} \rho_{n-1}\sin\left(\theta_{n-1} \right) \nonumber \\
    && - c k_{n-1} \rho_{n-1} \rho_{n-2} \sin\left(\theta_{n-2}\right) \nonumber \\
    && + \rho^f_n \cos(\phi^f_n - \phi_n) + D_n \rho_n\,, \label{eq:rho}
\end{eqnarray}
whereas the triad phase evolution acquires more terms since it is a linear combination of the evolution of three phases, and becomes
\begin{eqnarray}
    && \frac{d\theta_n}{dt} = k_{n+3} a \frac{\rho_{n+3}\rho_{n+4}}{\rho_{n+2}} \cos\left(\theta_{n+2}\right) \nonumber \\
    && + k_{n+2} \rho_{n+3}\left(b \frac{\rho_{n+1}}{\rho_{n+2}} - a \frac{\rho_{n+2}}{\rho_{n+1}} \right)\cos\left(\theta_{n+1}\right) \nonumber \\
    && - k_{n+1} \left(c\frac{\rho_{n}\rho_{n+1}}{\rho_{n+2}} + b \frac{\rho_{n}\rho_{n+2}}{\rho_{n+1}} + a \frac{\rho_{n+1}\rho_{n+2}}{\rho_{n}}\right)\cos\left(\theta_{n}\right) \nonumber \\
    && + k_n \rho_{n-1} \left(c\frac{\rho_{n}}{\rho_{n+1}} - b \frac{\rho_{n+1}}{\rho_{n}} \right)\cos\left(\theta_{n-1}\right) \nonumber \\
    && + k_{n-1} c \frac{\rho_{n-2}\rho_{n-1}}{\rho_{n}} \cos\left(\theta_{n-2}\right) \nonumber \\
    && + \frac{\rho^f_{n+2}}{\rho_{n+2}} \sin(\phi^f_{n+2} - \phi_{n+2}) - \frac{\rho^f_{n+1}}{\rho_{n+1}} \sin(\phi^f_{n+1} - \phi_{n+1}) \nonumber \\
    && - \frac{\rho^f_{n}}{\rho_{n}} \sin(\phi^f_{n} - \phi_{n}) \,. \label{eq:theta}
\end{eqnarray}

Since only the 2D-like shell models have a second sign-definite quadratic invariant, in this study we will consider only the flux of energy across scales. Due to the local nature of the equations, the energy flux through shell $n$ can be expressed simply as a sum of two terms:
\begin{eqnarray}
    \Pi_n &=&   \frac{1}{2} \sum_{m=0}^{n} \frac{d}{ dt} |u_m|^2  \label{eq:flux_full} \\
    &=& k_n \mathcal{I}\left[\lambda a u_{n+2}u^*_{n+1}u^*_n + (a + b) u_{n+1} u^*_{n} u^*_{n-1} \right] \nonumber \\
    &=& k_n \rho_n \rho_{n+1} \left[ \lambda a \rho_{n+2} \sin \left(\theta_n\right)+ (a + b) \rho_{n-1} \sin \left( \theta_{n-1}\right) \right], \nonumber
\end{eqnarray}
where $\mathcal{I}[\cdot]$ represents the imaginary component. Although we will not be discussing the flux of positive-definite $H$, we show its expression for completeness. Choosing the bottom sign in Eq. (\ref{eq:Ab_def}) so that $b =-\lambda^{-\gamma}-1$, the flux of the positive-definite $H$ is
\begin{eqnarray}
     \Pi^H_n &=& \frac{1}{2} \sum_{m=0}^{n} \frac{d}{ dt} k_m^\gamma |u_m|^2 \label{eq:flux_H_full}\\
     &=& k^{\gamma+1}_n \mathcal{I}\left[\lambda u_{n+2}u^*_{n+1}u^*_n - u_{n+1} u^*_{n} u^*_{n-1} \right] \nonumber \\
    &=& k^{\gamma+1}_n \rho_n \rho_{n+1} \left[\lambda \rho_{n+2} \sin \left(\theta_n\right) - \rho_{n-1} \sin \left( \theta_{n-1}\right) \right]. \nonumber
\end{eqnarray}
    
\subsection{Phase-only model \label{subsec:phase_only}}
Up to now, we have described the `full' shell model, where both the complex magnitude $\rho_n$ and the triad phase $\theta_n$ evolve in time. In this work we focus our attention on the triad phase dynamics and how they determine the flux of conserved quantities across inertial ranges with power-law energy spectra. For this reason, we will work with a \textit{phase-only} version of the shell model described above, where we fix the magnitude at each mode and evolve only the dynamics of the triad phases. This not only makes the description of the dynamics simpler, but also allows for a better control over the range of parameters and behaviors studied. 
The phase-only formulation has already been implemented in past work on the 1D Burgers \cite{Murray2018,Arguedas2022}, where it has revealed rich dynamics and insights for the full 1D Burgers system. 

As in previous work, and mimicking an inertial range with a power-law energy spectra, we will fix the magnitude of each mode:
\begin{equation}
    \rho_n = \rho_0 (k_n/k_0)^{\alpha/2} = \rho_0 \lambda^{\alpha n/2}\,, \label{eq:rho_def}
\end{equation}
resulting in an energy spectrum with exponent $\alpha$: $E_n = |u_n|^2/2 = \rho_0^2 \lambda^{\alpha n} / 2$. The exponents $\alpha$ and $\gamma$ are the control parameters in our study of the triad phase dynamics. Because the magnitudes do not evolve, energy is conserved, and so we will ignore viscosity when considering the phase-only model. For the sake of simplicity, we will ignore forcing as well (see \cite{Moradi2017} for a study of the effect of forcing on the phases in 1D Burgers). By plugging equations (\ref{eq:Ab_def}) and (\ref{eq:rho_def}) into Eq. (\ref{eq:theta}) and simplifying, we arrive at the dynamical equation for the triad phase, 
\begin{eqnarray}
    && \frac{d\theta_n}{dt} = k_0 \rho_0 \lambda^{n(1+\alpha/2)} \left[ \lambda^{(3+5\alpha/2)}\cos\left(\theta_{n+2}\right) \right. \nonumber \\
    && + \left(\pm \lambda^{2+\alpha-\gamma} - \lambda^{2+\alpha} - \lambda^{2+2\alpha}\right)\cos\left(\theta_{n+1}\right) \nonumber \\
    && - \left(\mp\lambda^{1-\alpha/2-\gamma} -\lambda^{1+\alpha/2}\pm\lambda^{1+\alpha/2-\gamma} + \lambda^{1+3\alpha/2}\right)\cos\left(\theta_{n}\right) \nonumber \\
    && + \left(\mp\lambda^{-\alpha-\gamma} \mp \lambda^{-\gamma} + 1\right)\cos\left(\theta_{n-1}\right) \nonumber \\
    && \mp \left. \lambda^{-1-3\alpha/2-\gamma} \cos\left(\theta_{n-2}\right)\right] \,. \label{eq:theta_powerlaw_gamma}
\end{eqnarray}
We notice here that the explicit $n$-dependence scales out from each term, resulting in a coefficient $k_0 \rho_0 \lambda^{n(1+\alpha/2)}$ that multiplies all terms. We define a new dynamical time-scale for each mode based on this quantity, namely $t^*_n \coloneq k_0 \rho_0 \lambda^{n(1+\alpha/2)} t$. Doing so gives us our final phase-only equation for the triad phase,
\begin{eqnarray}
    && \frac{d\theta_n}{dt_n^*} =  \sum_{m=n-2}^{n+2}K_{mn} \cos(\theta_m) \nonumber \\
    &&  =  \lambda^{(3+5\alpha/2)}\cos\left(\theta_{n+2}\right) \nonumber \\
    && + \left(\pm \lambda^{2+\alpha-\gamma} - \lambda^{2+\alpha} - \lambda^{2+2\alpha}\right)\cos\left(\theta_{n+1}\right) \nonumber \\
    && - \left(\mp\lambda^{1-\alpha/2-\gamma} -\lambda^{1+\alpha/2}\pm\lambda^{1+\alpha/2-\gamma} + \lambda^{1+3\alpha/2}\right)\cos\left(\theta_{n}\right) \nonumber \\
    && + \left(\mp\lambda^{-\alpha-\gamma} \mp \lambda^{-\gamma} + 1\right)\cos\left(\theta_{n-1}\right) \nonumber \\
    && \mp \lambda^{-1-3\alpha/2-\gamma} \cos\left(\theta_{n-2}\right) \,. \label{eq:theta_phase_only_final}
\end{eqnarray}
In the first line, we introduced a shorthand for each of the coefficients which will be used later. The lack of explicit $n$-dependence of the prefactors on the right hand side of Eq. (\ref{eq:theta_phase_only_final}) indicates that the system admits a steady-state solution that is independent of $n$. This will result in steady-state statistics of $\theta_n$ that do not depend on $n$.

Although no energy is being transferred across scales in the phase-only model, we can still measure the energy flux that would occur in the full shell model, given a power law inertial range with spectral exponent $\alpha$ and $n$-independent triad phase statistics. 
Plugging in Eq. (\ref{eq:rho_def}) and the assumption of $n$-independence for $\langle \sin(\theta_n)\rangle$ into the expression for energy flux, Eq. (\ref{eq:flux_full}), we get
\begin{eqnarray}
     \frac{\langle \Pi_n \rangle}{k_n \rho_n^3} &=& \frac{\langle \Pi_n \rangle}{k_0 \rho_0^3 \lambda^{n(1+3\alpha/2)}} \nonumber \\
     &\approx& \left(\lambda^{1+3\alpha/2} \pm \lambda^{-\gamma}\right)\langle \sin (\theta) \rangle. \label{eq:flux_en_final}
\end{eqnarray}
We have removed the $n$ subscript in $\langle \sin(\theta)\rangle$ since we have assumed it has no $n$-dependence. We will call the value of $\langle \sin(\theta) \rangle$ the \textit{alignment strength}, since it measures how aligned the triad phase $\theta$ is to the `maximum flux' value $\pi/2$ (incidentally, it also measures a correlation between the phases of each individual mode in the triad). Its statistics will be determined by the dynamics of $\theta$, and will therefore depend on the control parameters $\alpha$ and $\gamma$. Predicting its behavior is one of the main goals of our study because it is a main component of the would-be energy flux.

Although the individual terms on the left hand side of equation (\ref{eq:flux_en_final}) are $n$-dependent, the right hand side is not. This makes the ratio $\langle \Pi_n \rangle / (k_n\rho_n^3)$ an ideal measure of the energy flux for each phase-only shell model run, since it is $n$-independent and can be averaged in $n$ for better statistical convergence. Its sign will correspond to the sign of the energy flux, but notice that the flux itself $\langle \Pi_n \rangle$ can depend on $n$. Indeed, under our assumption of $n$-independent triad phase statistics, the \textit{only} $n$-independent flux (a requirement of a true inertial range of the full model) occurs when $\alpha = -2/3$. It corresponds to the Kolmogorov solution for the shell models (with a freedom to choose the sign of $\langle \sin(\theta)\rangle = \pm 1$). Another thing to note from expression (\ref{eq:flux_en_final}) is that the energy flux is determined by both $\langle \sin(\theta) \rangle$ and a coefficient that comes from the magnitudes $\rho_n$ and parameter $b$, giving an explicit dependence on $\alpha$ and $\gamma$, respectively. 

As is evident by Eq. (\ref{eq:theta_phase_only_final}), by fixing the shell magnitudes we explicitly ignore the contributions from their evolution and fluctuations on the dynamics of the triad phase. We also implicitly assume that correlations between fluctuating components of $\rho_m$ and $\theta_n$ are negligible for any pair of shells $m$ and $n$. Despite these strong assumptions, we found that the comparison of the statistics of $\theta_n$ between full 3D-like shell models and their phase-only counterpart (with the same spectral exponent) agree quite well, at least when considering the steady state probability distribution and $n$-independence. The two models will likely not agree when considering higher-order statistical measures, such as intermittency, but that is beyond the scope of this paper. As for 2D-like shell models, the power-law inertial range assumption and $n$-independence seem to break down in the full shell model (Appendix \ref{app:2D}). The reason for this is uncovered in our work on triad phase dynamics in the next section, and its implications will be discussed in the conclusion section.

\section{Simplified triad phase dynamics: noisy phase oscillator \label{sec:theory}}
Although many simplifications have already been made to arrive at a phase-only shell model, the evolution of $\theta_n$ remains coupled to that of other neighboring modes through nonlinear terms. This results in chaotic nonlinear dynamics, making it difficult to make any predictions for the statistics of $\theta_n$, which is our ultimate goal. Further simplifications must be made to facilitate analysis. 

Our key simplifying assumption for this work is the following: we will treat the sum of all neighboring triads in Eq. (\ref{eq:theta_phase_only_final}) as a single \textit{noise} variable $\xi_n$, in other words $\sum_{m\neq n} K_{mn} \cos(\theta_m) \mapsto \xi_n$.
In doing so, we implicitly neglect any statistical dependence between pairs of triad phases $\theta_n, \theta_m$ for all $n \neq m$, and invoke the central limit theorem to ensure that $\xi_n$ is a Gaussian variable. Namely, our key assumption neglects the phenomenon of triad-phase synchronization, whereby triad phases corresponding to different scales are correlated in time so as to enhance the flux. Such a phenomenon is observed in the 1D Burgers equation \cite{Murray2018} and in shell models \cite{CarrollThesis,Manfredini2025}, and is directly related to extreme events and intermittency. Statistical independence of the triad phases was first proposed as a simplifying assumption in a cascade model of turbulence, with a focus on intermittency \cite{Benzi1993}. However, shell model simulations show that this is not entirely true. Indeed, this cannot hold if intermittency is to be present in the system, since the only constant flux solution for independent triad phases, as mentioned in the previous section, is the self-similar Kolmogorov solution \cite{Eyink2003}.  
Although our key assumption may not precisely hold, we show here that correlations between neighboring triad phases are weak enough that our model does indeed provide a sufficiently accurate description of the steady state statistics of $\theta_n$ and its implications on the flux, which are our main goals for this work. In summary, our results should be valid in a regime where extreme events and intermittency do not contribute significantly to the statistics of the triad phases.

After making the substitution for the noise, our model equation for $\theta_n$ becomes
\begin{equation}
    \frac{d \theta_n}{dt^*_n} = \KK(\alpha,\gamma) \cos(\theta_n) + \xi_n, \label{eq:adler}
\end{equation}
where we have introduced the shorthand
\begin{eqnarray}
    && \KK(\alpha,\gamma) \coloneq K_{nn}(\alpha,\gamma) \label{eq:scriptK_def} \\ 
    &&= - \left(\mp\lambda^{1-\alpha/2-\gamma} -\lambda^{1+\alpha/2}\pm\lambda^{1+\alpha/2-\gamma} + \lambda^{1+3\alpha/2}\right). \nonumber
\end{eqnarray}
Equation (\ref{eq:adler}) is that of a noisy phase oscillator, also known as the Adler equation (with an intrinsic frequency equal to zero).  The deterministic dynamics of the triad phase are thus solely determined by the so-called `self-interaction term', $\KK\cos(\theta_n)$. The simplicity of the deterministic Adler equation gives us direct access to the fixed points and their stability. The two fixed points of Eq. (\ref{eq:adler}) are $\pm \pi/2$, with $\pi/2$ being stable for $\KK>0$ whereas $-\pi/2$ is stable for $\KK<0$. This already hints at implications of the sign of $\KK$ on the sign of the alignment strength, which is $\sin(\theta)=\pm 1$ for the deterministic fixed point values of $\theta=\pm \pi/2$, as well as the sign of the flux.

Before moving on to the full implications of the nondeterministic dynamics of our model for the triad phase evolution, we investigate the validity of our assumptions using a pair of phase-only simulations in the 3D- and 2D-like regimes. Both runs have $N=28$ modes, $\lambda=1.5$, $k_0 = \lambda^{-14}$, $\rho_0 = 10^{-0.5} k_0^{\alpha/2}$, $a=1$, and were initialized with random phases. The 3D-like run has a value of $\alpha = -2/3$ and $\gamma=1$ (so that $b=-1/3$), whereas the 2D-like run has a value of $\alpha=-2$ and $\gamma=0.855$ (so that $b=-1.707$). Equation (\ref{eq:theta}) was integrated in time using a fourth order Runge-Kutta method with time step $\Delta t = 10^{-2}$. We test our assumptions by outputting the supposed `noise' variable $\widetilde{\xi}_n = \sum_{m\neq n} K_{mn} \cos(\theta_m)$ for a few modes $n = \{N/4, N/3, N/2\}$. Note that for our analytical considerations $\xi_n$ is a true Gaussian random variable, whereas $\widetilde{\xi}_n$ is deterministic, albeit chaotic. 
Our goal is to see how close $\widetilde{\xi}_n$ comes to being a true Gaussian random variable like $\xi_n$, and to investigate its properties for modeling purposes.

\begin{figure}
\includegraphics[width=0.75\textwidth]{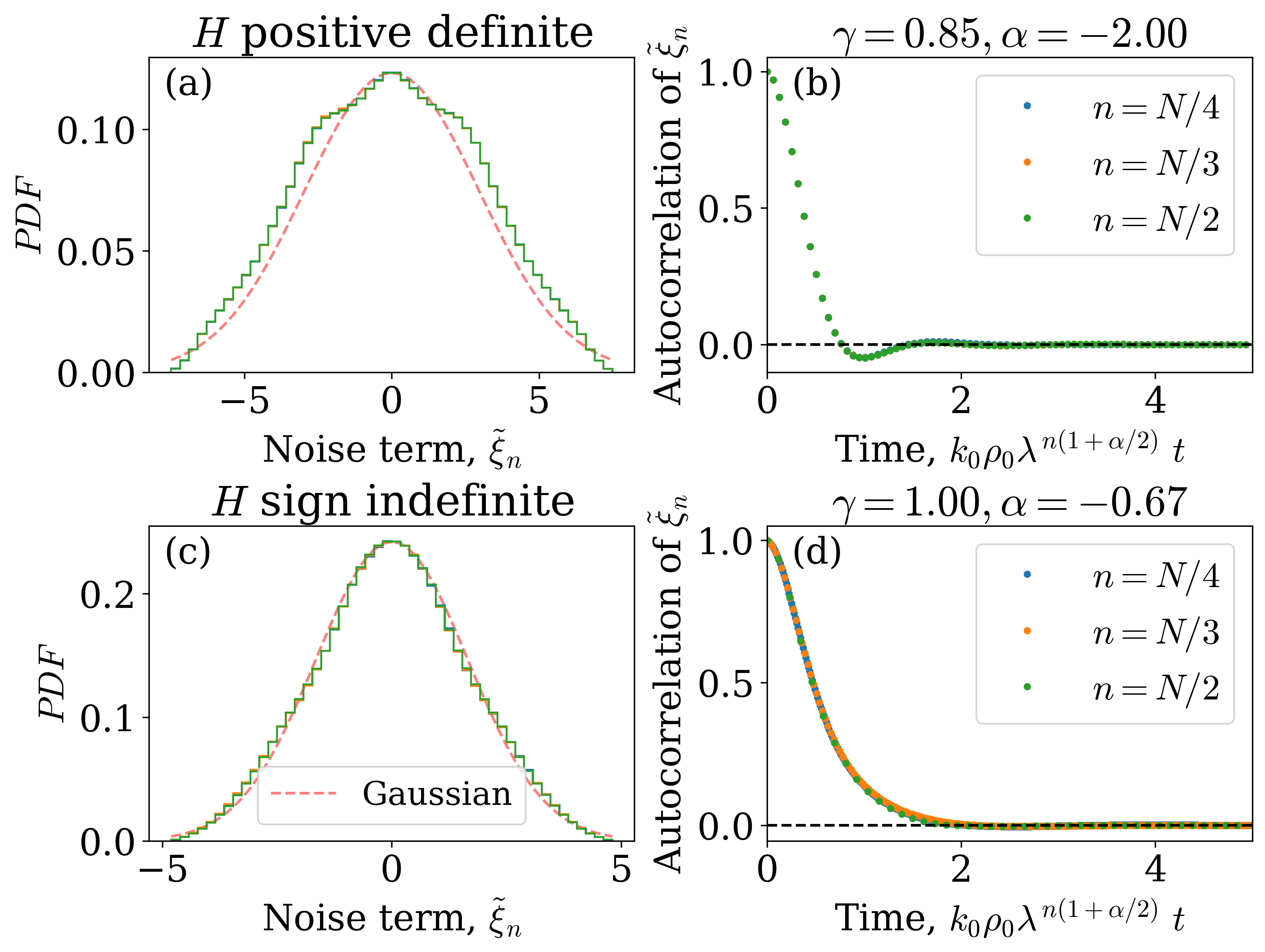}%
\caption{Statistics of $\widetilde{\xi}_n$, the `noise' variable measured from phase-only shell model runs of a 2D-like case (top row, (a) and (b)) and a 3D-like case (bottom row, (c) and (d)). The PDFs both show an approximately Gaussian shape(dashed red line), with a maximum at zero and a fall-off towards zero. Both autocorrelation functions have time-scales of order one. That of the 2D-like case, (b), shows a decaying and oscillating autocorrelation function, hinting at colored harmonic noise.
Each plot shows statistics for $n=N/4$ (blue), $n=N/3$ (orange), and $n=N/2$ (green). The overlap of these curves demonstrates the $n$-independent statistics of the re-scaled variables.   \label{fig:fig1}}
\end{figure}

The probability distribution function (PDF) and autocorrelation function for $\widetilde{\xi}_n$ are found in Figure \ref{fig:fig1} for both runs. Having divided all terms by $k_0\rho_0 \lambda^{n(1+\alpha/2)}$ and defining $t^*$, we see that, as expected, the steady-state statistics at each $n$ collapse onto one curve.
Despite being the sum of only four terms, the PDF of $\widetilde{\xi}_n$ looks approximately Gaussian (Figure \ref{fig:fig1}(a),(c)), unlike the PDFs of each individual $K_{mn}\cos(\theta_m)$ term. This will justify our analysis of the statistics of Eq. (\ref{eq:adler}) via the Fokker-Planck equation below. 
The autocorrelation function of the noise term $\widetilde{\xi}_n$ reveals an order-one time-scale for the dynamics (Figure \ref{fig:fig1}(b)(d)), which is the same time-scale as the dynamics of $\theta_n$. This suggests that our model noise $\xi_n$ should not be white noise. Furthermore, both autocorrelation functions are not exponential in nature, suggesting that modeling the noise as an Ornstein–Uhlenbeck process is not the most accurate representation. Instead, we found that colored harmonic noise gave the best match to our data \cite{SchimanskyGeier1990}.

Under a few reasonable assumptions, it is possible to compute the steady-state statistics of Eq. (\ref{eq:adler}) taking $\xi_n$ to be colored harmonic noise (see Appendix \ref{app:noise}). Indeed, this gives the best fits to the statistics of $\theta_n$.
Although more accurate, we found that the complexity of the resulting expressions for the statistics limited our ability to interpret and understand the results, and made the predictions of quantities of interest more difficult to compute.
Considering that our goal is to give a first-order description, we choose instead to model our noise using the simplest possible choice: white noise, such that $\overline{\xi_n(t) \xi_n(t+s) } = \DD \delta(s),$ where $\overline{\cdot}$ denotes an ensemble average and $\DD$ is the resulting `effective' noise amplitude which best matches the data.
Modeling the noise as white, in turn, reduces the undetermined free parameters in the model from three (for the colored harmonic noise) to one (for the white noise). This reduces the sensitivity of the fits to be performed in subsequent sections: whereas the colored harmonic noise fits are non-unique and depend strongly on initial guesses, the white noise fits for $\DD$ are robust. Even though $\widetilde{\xi}_n$ clearly has a similar time-scale as $\theta_n$, modeling $\xi_n$ as white noise will turn out to be sufficient to describe the steady-state statistics of $\theta_n$ to first order. In what follows we drop subscripts $n$, since the steady-state statistics do not depend on the mode $n$.

\begin{figure}
\includegraphics[width=0.65\textwidth]{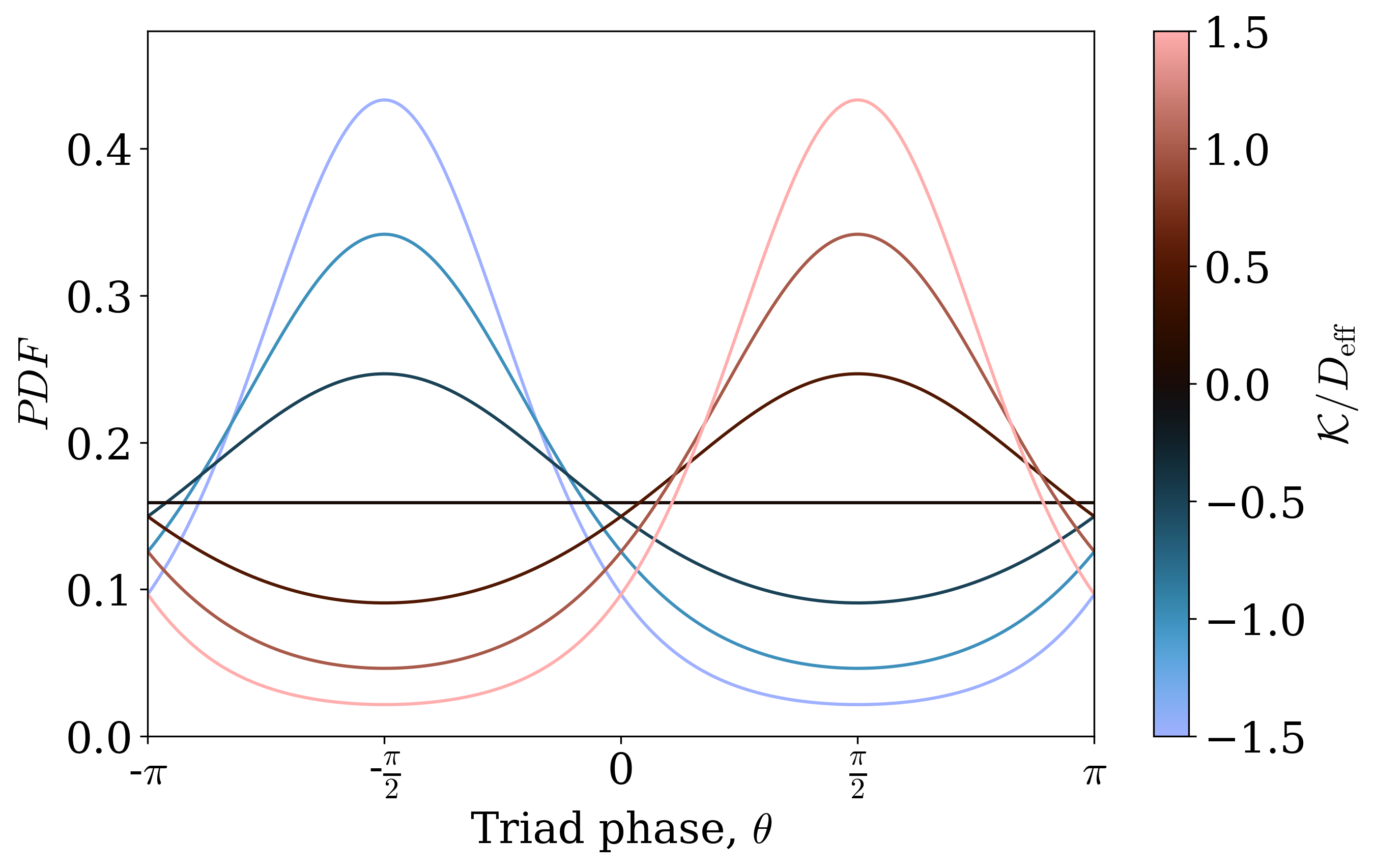}%
\caption{A plot of the predicted PDF of $\theta$, Eq. (\ref{eq:PDF_theta}), based on our model for the triad phase dynamics, Eq. (\ref{eq:adler}), assuming white noise. Multiple values of $\KK/\DD$ are shown, with red color corresponding to positive $\KK/\DD$, blue to negative $\KK/\DD$, and black to $\KK/\DD = 0$. Notice how the position of the maximum value of the PDF, $\pm \pi/2$, changes sign depending on the sign of $\KK/\DD$. \label{fig:fig2}}
\end{figure}

Given our assumptions on the noise $\xi$, it is straightforward to compute the Fokker-Planck equation associated with Eq. (\ref{eq:adler}) and solve for the steady-state PDF \cite{VanKampen1992}, giving
\begin{equation}
    PDF(\theta) = \frac{1}{\mathcal{Z}} \exp\left[\frac{\KK(\alpha,\gamma)}{\DD} \sin(\theta) \right], \label{eq:PDF_theta}
\end{equation}
where $\mathcal{Z}$ is the normalization factor, equal to $\mathcal{Z} = 2 \pi I_0[\KK/\DD]$, and $I_i[z]$ is the modified Bessel Function of the first kind \cite{Weisstein}. The resulting PDF, Eq. (\ref{eq:PDF_theta}), is seen in Figure \ref{fig:fig2} for various values of $\KK/\DD$, including different signs of $\KK$ ($\DD>0$ by definition). 
The most likely value of $\theta$ corresponds to the stable fixed point of the deterministic dynamics, and therefore depends on the sign of $\KK$. When $\KK=0$ the distribution is uniform, suggesting no triad-phase alignments and thus no correlation between individual phases. The PDF closely resembles that which has been observed in a previous study of a full 3D-like shell model, where the authors found the best fit to be of the same functional form, with $\KK/\DD = -1.2$ \cite{Eyink2003}.

Given the PDF of the triad phase, the alignment strength can be computed to be 
\begin{equation}
    \langle \sin (\theta) \rangle = I_1[\KK/\DD]/I_0[\KK/\DD]. \label{eq:sync_def}
\end{equation}
It has the property that the sign of $\langle \sin (\theta) \rangle$ is the same as that of $\KK$, and that $\langle \sin (\theta) \rangle = 0$ when $\KK=0$. With an expression for the alignment strength, we have a direct prediction for the flux of energy via Eqs. (\ref{eq:flux_en_final}), (\ref{eq:scriptK_def}) and (\ref{eq:sync_def}), which now depends explicitly on $\alpha, \gamma$, and an unknown $\DD$. We have therefore managed to directly link the dynamical equation of the triad phase to a prediction for the energy flux. Importantly, the \textit{direction} of the energy flux does not depend on the value of our free parameter, $\DD$, meaning that we have a prediction for the cascade direction based entirely on known control parameters. 
            
\section{Comparison with shell model simulations \label{sec:sims}}
Now that we have established our model for the statistics of $\theta$, we will compare our predictions to a suite of phase-only shell model simulations in both the 2D-like and 3D-like cases. All runs have $N=28$ modes, $\lambda=1.5$, $k_0 = \lambda^{-14}$, $\rho_0 = 10^{-0.5} k_0^{\alpha/2}$, $a=1$, and were initialized with random phases. Equation (\ref{eq:theta}) was integrated in time using a fourth order Runge-Kutta method with time step $\Delta t = 10^{-2}$. The python code used to perform these phase-only shell model simulations, along with example run directories, is available on GitHub \cite{Benavides2025Code}.

We performed a total of 600 phase-only shell model runs (312 for the 2D-like runs and 288 for the 3D-like runs), varying the spectral slope $\alpha$ and the exponent $\gamma$ of $k_n$ in the second conserved quantity $H$. For both sets of runs, we varied the spectral slope between $\alpha=-2$ and $\alpha=0$, with increments $\Delta \alpha = 0.087$ resulting in 24 different values of $\alpha$. For the 2D-like runs, the exponent of $k_n$ in $H$ was varied between $\gamma = 0.13$ (corresponding to $b = -1.95$) and $\gamma = 3.42$ (corresponding to $b = -1.25$), with a constant increment in $b$, $\Delta b = 0.05$, resulting in 13 different values of $\gamma$. For the 3D-like runs, $\gamma$ was varied between $\gamma=0.26$ (corresponding to $b=-0.1$) and $\gamma=2.97$ (corresponding to $b=-0.7$), with an approximately constant increment in $\gamma$, $\Delta \gamma \approx 0.25$, resulting in 12 values of $\gamma$.

For each run, we gathered statistics of the triad phase in the form of a PDF of $\theta_n$, and computed the time-averaged energy flux $\langle \Pi_n \rangle$, Eq. (\ref{eq:flux_full}), for all modes $n$. Since the PDF of $\theta_n$ and the ratio $\langle \Pi_n \rangle/(k_n \rho_n^3)$ do not depend on $n$, we then averaged these quantities in $n$ (away from $n=0$ and $n=N$ to avoid boundary effects) for better statistical convergence. Therefore, each run produced two $n$-averaged observables: the rescaled energy flux, $\langle \Pi_n \rangle/(k_n \rho_n^3)$, and a PDF for $\theta$. From the latter we can calculate the alignment strength, $\langle \sin(\theta) \rangle$, and estimate $\DD$ using a fit of the PDF with Eq. (\ref{eq:PDF_theta}).

\subsection{Positive-definite $H$ \label{subsec:2D_like}}
\begin{figure*}
\includegraphics[width=\textwidth]{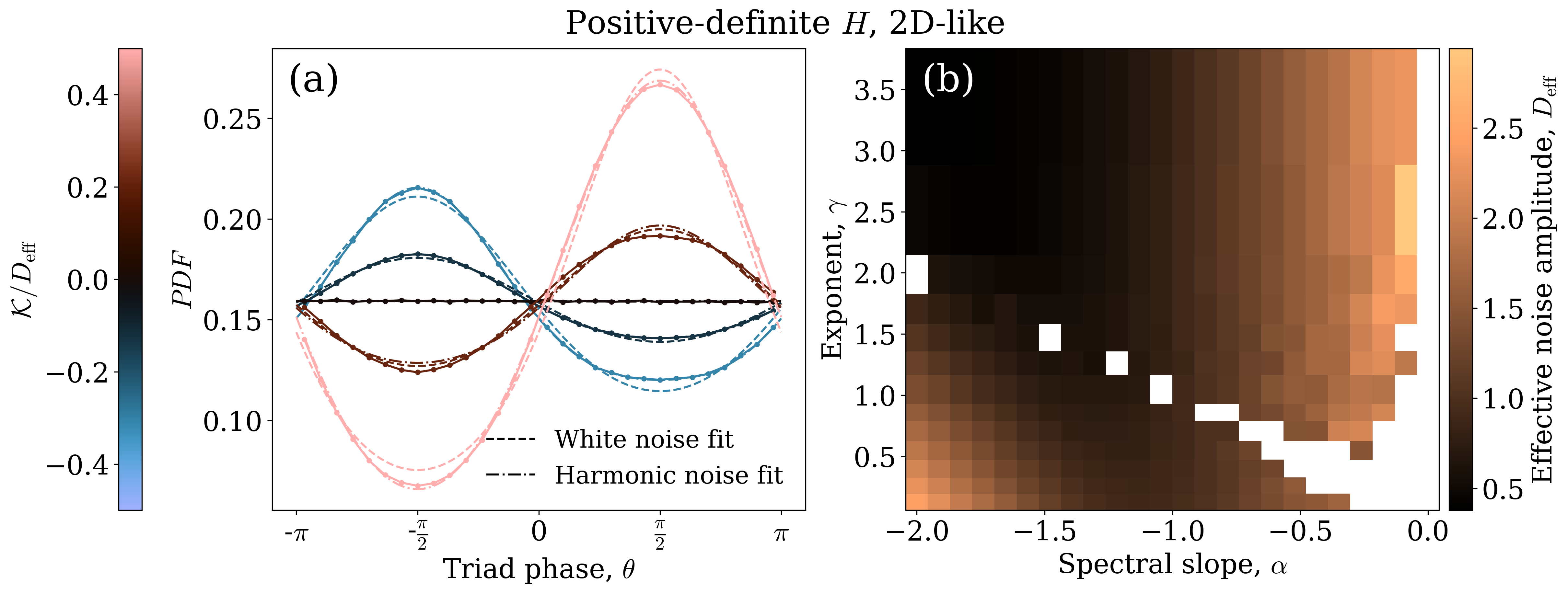}%
\caption{Measured PDFs, (a), and $\DD$ fits, (b), for our suite of 2D-like runs with positive-definite $H$. Panel (a) shows a demonstrative sample of PDFs of $\theta_n$, averaged in $n$, observed in our runs, with $\KK/\DD$ positive (red), negative (blue), and zero (black). PDF fits for the white noise model (dashed lines), based on Eq. (\ref{eq:PDF_theta}), and the colored harmonic noise model (dot-dashed lines), based on Eq. (\ref{eq:HN_PDF}), are also shown. Panel (b) shows the measured values of $\DD$ based on fits to Eq. (\ref{eq:PDF_theta}) for all values of spectral slope $\alpha$ and second invariant exponent $\gamma$. The masked data (in white) represents runs where the PDFs are sufficiently flat such that fitting for $\DD$ was not possible.  \label{fig:fig3}}
\end{figure*}

We first focus on the 2D-like runs, whose second conserved quantity $H$ is positive-definite. Figure \ref{fig:fig3}(a) shows PDFs of the triad phase for five different runs, chosen to demonstrate the variety of behavior found, including different alignment strengths, PDF maxima at $\pm \pi/2$, and an unsynchronized uniform distribution. 
These PDFs are fit with both noise models: the white noise fit is seen in a dashed line whereas the colored harmonic noise fit is seen in the dash-dot line. Particularly for the runs with $\KK<0$, the colored harmonic noise PDF does a remarkable job of fitting the observed PDFs--so much so that the fits are hard to see. That said, one can also see that the white noise PDF is not far off and, importantly, captures both the alignment strength and sign with only one fitting parameter $\DD$. 

The estimated value of $\DD$ for each run, based on the fit to the PDF of $\theta$, is shown in Figure \ref{fig:fig3}(b). The masked data (in white) represents runs where the PDFs are sufficiently flat such that fitting for $\DD$ was not possible.
The estimated values of $\DD$ are all order one in magnitude, with a mean of $\DD=1.15$ and extremes ranging from about $\DD\approx0.5$ to $\DD\approx3$. This suggests that our model for the triad phase statistics with a fixed $\DD$ is a good first-order approximation to the true statistics. Furthermore, and more importantly, none of the estimated values of $\DD$ are negative, meaning that $\KK$ does indeed determine the sign of the alignment strength, as it does in our model. Figure \ref{fig:fig3}(b) does show a clear next-order functional dependence of $\DD$ on $\alpha$ and $\gamma$. This likely comes from the $\alpha$ and $\gamma$ dependence of the $K_{mn}$ terms that make up the noise term $\widetilde{\xi}_n$, but we were not able to analytically compute any prediction for this dependence.

\begin{figure*}
\includegraphics[width=\textwidth]{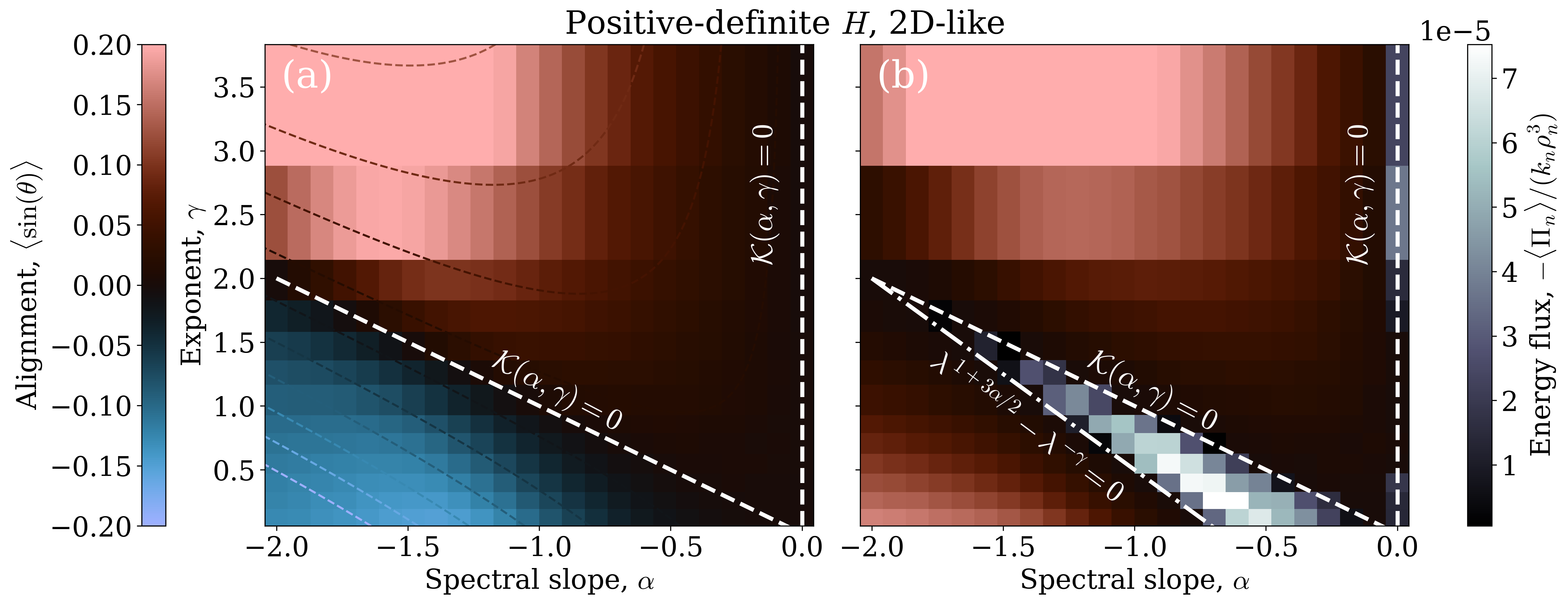}%
\caption{Time- and $n$-averaged alignment strength $\langle \sin(\theta)\rangle$, (a), and rescaled energy flux $\langle \Pi_n \rangle/(k_n \rho^3_n)$, (b), for our suite of 2D-like phase-only model runs. Panel (a) shows the alignment, which is both positive (red) and negative (blue). Lines of zero alignment predicted by our {analytical} model for the triad phase dynamics are seen as white dashed lines, corresponding to solutions to $\KK=0$. Panel (b) shows the rescaled energy flux, which can also be both positive (red) and negative (blue-white). we have re-scaled the negative flux to be able to distinguish it from zero. The colorbar scale for positive flux in (b) is $2.5 \times 10^{-3}$. The dashed white lines are the same zero-alignment lines in (a), whereas the dot-dashed line corresponds to a third zero-flux line, based on the coefficient for the flux in Eq. (\ref{eq:flux_en_final}). \label{fig:fig4}}
\end{figure*}

We turn now to the measured values of the alignment strength $\langle \sin(\theta) \rangle$, cf. Figure \ref{fig:fig4}(a). As hinted in Figure \ref{fig:fig3}(a), the alignment can be both positive and negative, with the latter occurring for steeper spectra and lower values of the exponent $\gamma$. This is seen in Figure \ref{fig:fig4}(a) as blue regions (negative alignment) and red regions (positive alignment). We now focus on the lines of zero alignment that are evident in this figure. For any value of $\gamma$, as the spectral exponent $\alpha$ approaches zero, the alignment strength goes to zero. In these cases, the individual phases are unsynchronized, resulting in zero net energy flux. Indeed, the flat energy spectrum at $\alpha = 0$ corresponds to the equipartition equilibrium solution for the full shell model.  There is another line along which the phases are not synchronized, which separates the positive and negative alignment regions in the explored parameter space.
Both lines of zero alignment can be explained with our analytical model for the statistics of the triad phase, which predicts zero alignment strength when $\KK = 0$. Solving $\KK(\alpha,\gamma) = 0$ choosing the bottom sign in Eq. (\ref{eq:scriptK_def}) results in two sets of solutions, represented by white dashed lines in Figure \ref{fig:fig4}(a): $\alpha_{c,1} = 0$, corresponding to the equipartition equilibrium solution mentioned above, and $\alpha_{c,2} = -\gamma$, corresponding to the line separating regions of positive and negative alignment. These zero-alignment lines agree remarkably well with the numerical results from the phase-only shell models. A more detailed comparison of the predicted alignment strength using Eq. (\ref{eq:sync_def}) with $\DD = 1.15$ (dashed colored contours, Figure \ref{fig:fig4}(a)), reveals a correct prediction of the sign of the alignment strength, and a correct prediction of the alignment-strength magnitudes which have, on average, a $32\%$ error (minimum error of $1 \%$, maximum error of $108 \%$ and median of $30\%$).

We turn now to the rescaled energy flux (Fig. \ref{fig:fig4}(b)), which shows, perhaps surprisingly, a larger tendency for \textit{forward} energy flux. Only a small region below the $\alpha_{c,2} = -\gamma$ line demonstrates an inverse flux. Not only is a forward energy flux more likely in this parameter space, it is also larger in magnitude than the inverse energy flux. Indeed, the red colorscale shown is for $0 < \langle \Pi_n \rangle / (k_0 \rho_0^3) < 2.5 \times 10^{-3}$, whereas that of the inverse cascade is more than an order of magnitude smaller (colorbar, Figure \ref{fig:fig4}(b)). This is a result of the fact that the small region where an inverse cascade appears coincides with weak triad phase alignment (Fig. \ref{fig:fig4}(a)). 
The behavior of the flux can be explained using Eq.~(\ref{eq:flux_en_final}) and our analytical model for the triad phase dynamics, Eq.~(\ref{eq:sync_def}). The energy flux changes sign due to both a change of sign of the alignment (at $\alpha_{c,2} = -\gamma$, white dashed line, Fig. \ref{fig:fig4}(b)) and a change of sign of the coefficient, which is not dependent on the behavior of the phases but instead simply on the energy spectrum. This occurs for $\alpha_{c,3} = -2(\gamma +1)/3$ (dot-dashed white line, Fig. \ref{fig:fig4}(b)). We see again that these lines agree remarkably well with zero measured flux of the phase-only shell model runs. 

We remind the reader that the measurement of what we call the energy flux, Eq.~(\ref{eq:flux_en_final}), is really a measure of the energy flux that would occur in the full shell model, given a power law inertial range with spectral exponent $\alpha$ and $n$-independent triad phase statistics. Therefore, our results do not suggest that full models with positive-definite $H$ will have a sustained forward energy cascade. Instead, they reveal which inertial range power law exponents support an inverse energy cascade. Take, for example, the Kolmogorov self-similar solution of $\alpha=-2/3$ at $\gamma = 1.26$. For an inverse cascading system, we must have negative alignment, meaning $\theta_n = -\pi/2$. With these values of $\alpha$ and $\gamma$, this solution sits above the $\KK=0$ line of $\alpha_{c,2}$ (Fig.~\ref{fig:fig4}(a)), and our model for the phase dynamics suggests that this fixed point is \textit{unstable}: instead, it is the $\theta_n = \pi/2$ fixed point that is stable. Therefore, the Kolmogorov self-similar solution for an inverse cascade range is not stable and the system must break either the assumption of $n$-independent triad phase statistics or the assumption of having $\alpha=-2/3$, or both. Indeed, a full model simulation of the system at $\gamma=1.26$ (Fig. \ref{fig:full_run_data}, Appendix \ref{app:2D}) reveals that both assumptions are broken: the system forms an inverse cascade range where $\langle \sin(\theta_n)\rangle$ varies with $n$ and which has an exponent $\alpha \approx -1.32$. This value of $\alpha$ now lies between $\alpha_{c,2}$ and $\alpha_{c,3}$ in the range where $\theta_n = -\pi/2$ is a stable fixed point of the model dynamics. In fact, Figure \ref{fig:full_alphas} in Appendix \ref{app:2D} shows that the same is true for all full runs: the final exponents of the inverse cascading inertial range all lie within the regions where negative flux is possible, based on our model's prediction. This is rather surprising, given the fact that our zero-flux lines are based on the assumption that the statistics are $n$-independent, yet all of the full runs have clearly $n$-dependent phase alignment. The breakdown of the $n$-independence of triad phase statistics is necessary for the presence of an energy flux that is constant in $n$ (Section \ref{subsec:phase_only}, Fig. \ref{fig:full_run_data}). 
Many possible energy spectra and triad phase distributions can be consistent with a particular constant flux solution. As we have seen above, considering the triad phase dynamics therefore acts to select the possible dynamical stable configuration that would be found in a full shell model simulation.

These considerations reveal an important implication for the shell model analogue of 2D turbulence, with $\gamma = 2$. Note that $\gamma = 2$ is the point where zero-flux lines $\alpha_{c,2}$ and $\alpha_{c,3}$ cross.  For this particular value of $\gamma$, $\theta_n = -\pi/2$ is an unstable fixed point for all values of $\alpha <0$. This suggests that no inverse cascading, power-law inertial range can form in this system. Our model for the triad phase dynamics thus offers another explanation for the observed behavior of shell models of 2D turbulence, where no inverse cascade was observed and a quasi-equilibrium solution formed instead \cite{Aurell1994,Ditlevsen1996,Gilbert2002,Gurcan2016}.

\subsection{Sign-indefinite $H$ \label{subsec:3D_like}}
\begin{figure*}
\includegraphics[width=\textwidth]{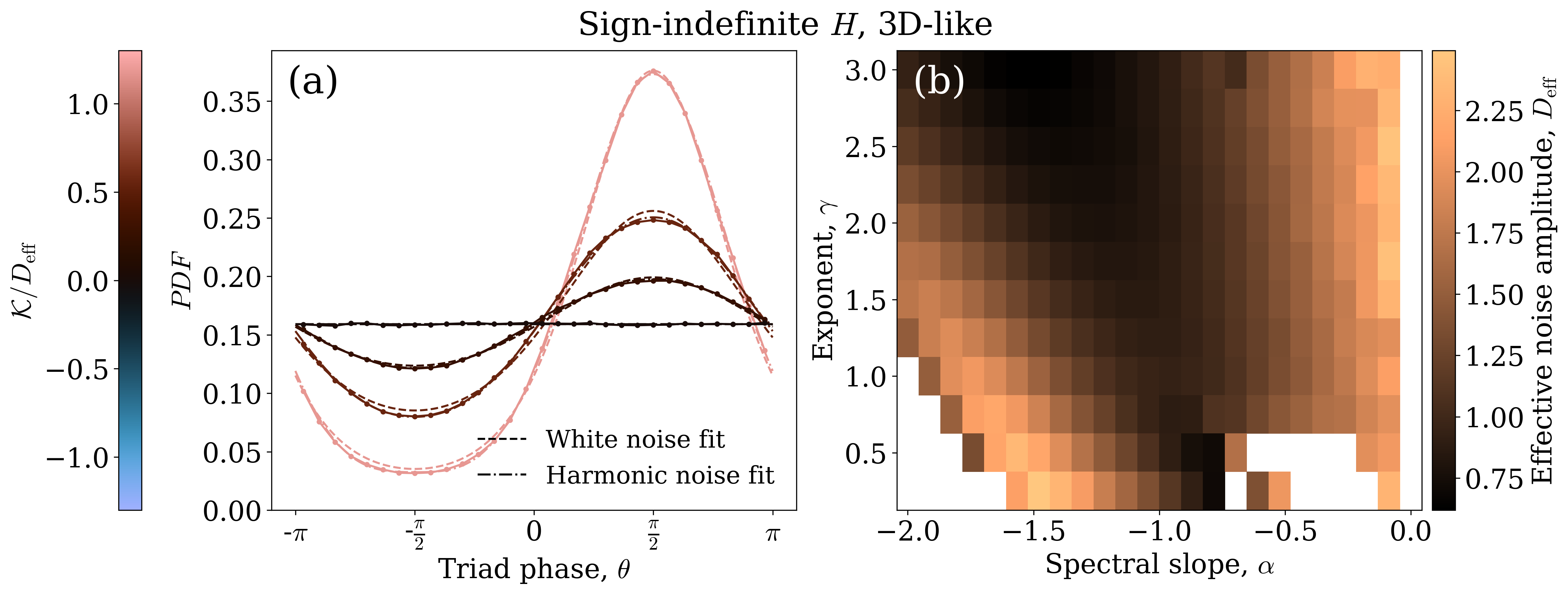}%
\caption{Measured PDFs, (a), and $\DD$ fits, (b), for our suite of 3D-like runs with sign-indefinite $H$. Panel (a) shows a demonstrative sample of PDFs of $\theta_n$, averaged in $n$, observed in our runs, with $\KK/\DD$ only being positive (red) and zero (black). PDF fits for the white noise model (dashed lines), based on Eq. (\ref{eq:PDF_theta}), and the colored harmonic noise model (dot-dashed lines), based on Eq. (\ref{eq:HN_PDF}), are also shown. Panel (b) shows the measured values of $\DD$ based on fits to Eq. (\ref{eq:PDF_theta}) for all values of spectral slope $\alpha$ and second invariant exponent $\gamma$. The masked data (in white) represents excluded runs with either flat PDFs ($\alpha=0$) or with non-chaotic dynamics (found for low values of $\gamma$). \label{fig:fig5}}
\end{figure*}

We perform the same analysis for the 3D-like runs, whose second conserved quantity $H$ is sign-indefinite (choosing the top sign in Eq.~(\ref{eq:Ab_def}) and subsequent definitions). Figure \ref{fig:fig5}(a) shows four PDFs of the triad phase, with varying alignment strength. As before, the colored harmonic noise fits are better and hardly visible, but the white noise model fits do a good job of representing the PDFs with only a single free parameter. The effective noise amplitude $\DD$ measured from fits of the PDFs of the triad phase is shown in Fig.~\ref{fig:fig5}(b). Masked data (in white) represents excluded runs with either flat PDFs ($\alpha=0$) or with non-chaotic dynamics (found for low values of $\gamma$). 
All of the measured $\DD$ values are positive and the average value among the runs is $\DD\approx 1.3$, with a minimum of $\DD=0.6$ and maximum of $\DD=2.5$. This suggests that the assumption of a constant $\DD$ of order one in our model is reasonable for the 3D-like runs. That said, like in the 2D-like runs, there is a second-order functional dependence of $\DD$ on $\alpha$ and $\gamma$ that our model cannot explain.

Notice that we show no PDFs whose maximum likelihood is at $\theta = -\pi/2$ (Fig.~\ref{fig:fig5}(a)). This is because, unlike the 2D-like runs, the 3D-like runs all have positive alignment, $\langle \sin(\theta)\rangle \geq 0$, for $\alpha < 0$. We can predict this analytically using our model for the triad phase dynamics, since the sign of $\langle\sin(\theta)\rangle$ is the same as that of $\KK$. Considering the definition of $\KK$, Eq.~(\ref{eq:scriptK_def}), we first note that $\alpha = 0$ is the only line along which $\KK=0$. It then remains to prove that $\partial_\alpha \KK|_{\alpha=0} < 0$ to show that $\KK > 0$ for $\alpha < 0$. Explicitly evaluating this derivative gives $\partial_\alpha \KK|_{\alpha=0} = -\log(\lambda) \lambda^{1-\gamma}(\lambda^\gamma+1)$. Since $\lambda>1$, we clearly see that this expression is indeed negative, proving our claim. It then follows that the energy flux is also positive for $\alpha < 0$, since the coefficient in front of $\langle \sin(\theta)\rangle$ in Eq.~(\ref{eq:flux_en_final}) is positive, given the choice of the top sign for the 3D-like case. Phase-only shell model runs confirm our prediction of positive alignment and energy flux for all runs. 

This result for the 3D-like case reveals why full shell model runs behave much closer to what is expected, with a positive energy flux and an inertial range with a power-law exponent close to the Kolmogorov self-similar solution of $\alpha = -2/3$. This power law solution with $\theta_n = \pi/2$, resulting in a forward energy cascade, is stable with respect to the triad phase dynamics and is therefore a viable inertial range. Indeed, in a full shell model simulation of the 3D-like case, with $\gamma=1$, we observe an inertial range with a power law exponent close to $\alpha=-2/3$ and $n$-independent triad phase statistics.

\section{Conclusions \label{sec:conclusions}}

In this work, we have investigated the dynamics of triad phases in hydrodynamic shell models. In both the Navier Stokes equations and shell models, triad phases play a crucial role in determining the direction and strength of the flux of quadratic conserved quantities across scales. By assuming that the terms containing neighboring triad phases can be treated as a noise, we modeled the triad phase dynamics of a phase-only shell model using the noisy Adler equation.
We then analytically solved the Fokker-Planck equation associated with this model, resulting in a prediction for the alignment strength, and therefore energy flux, of our system. Importantly, we found that the sign of the alignment is controlled solely by the sign of the coefficient of the `self-interaction term' in the triad phase dynamical equation, $\KK$ in Equations (\ref{eq:adler}) and (\ref{eq:scriptK_def}). This provides simple analytical predictions for the sign of the alignment and energy flux. We were able to show that the triad phase dynamics prevent the 2D turbulence shell model analogue (with $\gamma=2$) from sustaining an inverse-cascading power law inertial range, and proved that the 3D-like shell models all have positive alignment and energy flux. Our assumptions and predictions were confirmed by a suite of phase-only shell model simulations. 

In our analytical model for the triad phase dynamics, the sign of the alignment is determined by the linear stability properties of the two fixed points in the deterministic dynamics of an \textit{isolated} triad phase. This shares a striking resemblance to the `instability assumption' of Waleffe \cite{Waleffe1992,Waleffe1993}, whose predictions for the flux direction were also based on the stability properties of a single triad, and matched observations from simulations. Although Waleffe had little justification for this assumption \cite{Moffatt2014}, our work might offer some insight as to why it has correctly predicted observations: at least in shell models, correlations between neighboring triad phases are sufficiently weak so that the order-one behavior of a single triad is largely independent of its neighbors. Unlike Waleffe's instability assumption, our work goes beyond predicting the direction of the flux by directly producing an expected value for the alignment and flux.

The surprisingly simple description of the triad phase dynamics, and its direct prediction of the alignment and energy flux, begs the question of its extension to more realistic contexts and flows. Although we have seen that the conditions of scale-independent triad phase statistics approximately hold in the inertial ranges of full 3D-like shell model simulations, this is not the case in 2D-like full model simulations, as we have already noted in the main text and in Appendix \ref{app:2D}. This is due to the restriction in the inertial ranges of full systems requiring a constant-in-scale energy flux. That said, our simplified analytical model of the triad phase dynamics may still hold without having to invoke a power law energy spectrum and $n$-independent triad phase dynamics -- predictions of the triad phase behavior could depend on $n$.  
Also, an investigation of how the model's noise properties change between full and phase-only shell model runs is worth considering. Preliminary results from the two full shell model runs mentioned in this work reveal hints that the noise has a larger variance and larger tails in its PDF, but that otherwise the same properties hold. 

There is also the question of how our results translate to the Navier Stokes equations and what they might reveal about energy cascades in such systems. A forthcoming manuscript will address an extension of our approach to the two-dimensional Navier Stokes equation. However, the case of three-dimensional Navier Stokes is more complex. This is due to the presence of two complex velocity amplitudes (e.g. in the helical decomposition), which implies the presence of more than one type of triad phase for each triad. The interaction between triad phases of different types for a single triad may be nontrivial, resulting in a more complex representation of the triad phase dynamics.

A model of triad phase dynamics and their implication on the flux of conserved quantities for realistic flows could improve our understanding of established results in turbulence, but could also be helpful in understanding the behavior that deviates from what is expected. In the context of anisotropic turbulent flows, such as in geophysical and astrophysical systems, the energy can cascade both up- and down-scale in what is known as a bidirectional energy cascade \cite{Cho2001,Sorriso-Valvo2007,AlexakisReview,Qiu2022,Balwada2022,Marino2025}. The bidirectional cascade exhibits striking behavior as the relevant geophysical parameter (which produces anisotropy) is varied: starting from a fully forward cascade, the bidirectional cascade seems to appear via a bifurcation at a precise parameter value \cite{Seshasayanan2014,Seshasayanan2016,Benavides2017,Pestana2019,AlexakisReview,VanKan2020,vanKan2024}. Neither this bifurcation nor the functional dependence of the inverse-to-forward cascade ratio with the control parameter are well understood. We expect that similar triad phase dynamics models can shed light on the dynamical and physical mechanisms responsible for this behavior.

\begin{acknowledgments}
The authors would like to thank Alexandros Alexakis, Luca Biferale, Michele Buzzicotti, Javier Jim{\'e}nez and Adrian van Kan for helpful discussions during the development of this project and manuscript.
S.J.B. has received funding from the European Union’s Horizon 2022 research and innovation programme under the Marie Sk{\l}odowska-Curie grant agreement No. 101109237.

The code used to run the shell model simulations described in the main text is v1.0Z of `MHD shell model' \cite{Benavides2025Code}. It is preserved at \url{https://doi.org/10.5281/zenodo.15800546}, available via the Creative Commons Attribution 4.0 International (CC-BY) license. 

The Figshare repository \cite{Benavides2025Data}, preserved at \url{https://doi.org/10.6084/m9.figshare.29469659}, contains (i) full data sets for shell model runs; (ii) temporal data for the two runs used in Figure \ref{fig:fig1}; (iii) the scripts used to generate the figures in the main text.
\end{acknowledgments}

\appendix
\section{Colored harmonic noise \label{app:noise}}
In this Appendix, we describe the statistics of our triad phase model, Eq. (\ref{eq:adler}), assuming that $\xi_n$ is modeled as colored harmonic noise. These results are based on analysis performed by Schimansky-Geier \& Z{\"u}licke \cite{SchimanskyGeier1990}. The model for the noise variable is the following:
\begin{equation}
    \frac{d \xi}{dt} = y, \quad \frac{d y}{dt} = -\Gamma y - \Omega_0^2 \,\xi + \sqrt{2 D} \Omega_0^2\, \xi_W, \label{eq:HN}
\end{equation}
where $D$, $\Gamma$, and $\Omega_0$ are the free model parameters, and $\xi_W$ is a white noise with $\overline{\xi_W(t) \xi_W(t+s)} = \delta(s)$, where $\overline{\cdot}$ denotes ensemble average and $\delta$ denotes the Dirac delta distribution. For simplicity of exposition we removed the asterisk from the nondimensional time and the $n$ subscript from the variables in this Appendix. The autocorrelation function associated with this system is
\begin{equation}
    \overline{\xi(t) \xi(t+s)} = \frac{D \Omega_0^2}{\Gamma} \mathrm{e}^{-\Gamma s/2}\left[ \cos(\omega_1 s) + \frac{\Gamma}{2 \omega_1}\sin(\omega_1 s)\right],
\end{equation}
where $\omega_1= \sqrt{\Omega_0^2 - \Gamma^2/4}$.

After a change of variables, and a pair of assumptions that restrict our analysis to long time-scales \cite{SchimanskyGeier1990} (justified by our interest in steady state statistics), the resulting effective equation for the triad phase becomes:
\begin{equation}
    \frac{d \theta}{dt} = \frac{\KK \cos(\theta) + \sqrt{2D}\xi_W}{1+\tau \KK \sin(\theta) - \frac{\KK(\Gamma \sin(\theta) + \KK)}{(\Gamma+\KK\sin(\theta))}},
\end{equation}
where $\tau\coloneq \Gamma/\Omega_0^2$. The steady-state PDF for these dynamics can be found using the Fokker-Planck equation, resulting in
\begin{widetext}
\begin{eqnarray}
    PDF(\theta) &=& \frac{1}{\mathcal{Z}}\left[1+\tau \KK \sin(\theta) - \frac{\KK(\Gamma \sin(\theta) + \KK)}{(\Gamma+\KK\sin(\theta))}\right] \nonumber \\
    && \times \exp{\left[\frac{\KK}{D} \sin(\theta) - \frac{\tau \KK^2}{2 D}\cos(\theta)^2 - \frac{\Gamma^2-\KK^2}{D(\Gamma+\KK \sin(\theta))} - \frac{\Gamma}{D}\log(\Gamma + \KK \sin(\theta))\right]}. \label{eq:HN_PDF}
\end{eqnarray}
\end{widetext}
Equation (\ref{eq:HN_PDF}) was used to make the harmonic noise fits of the PDFs seen in Figures \ref{fig:fig3} and \ref{fig:fig5}.

\section{Comparison to 2D-like `full' shell model runs \label{app:2D}} 

In this Appendix, we briefly discuss 2D-like shell model runs of the non-phase-only (i.e., `full') shell model, Eq. \eqref{eq:full_eq}. We performed 11 runs varying $\gamma$ from $\gamma = 0.13$ to $\gamma = 3.4$. All runs have $N=35$ modes, $\lambda=1.5$, $k_0 = \lambda^{-15}$, and $a=1$. Both small-scale and large-scale dissipation were included, so that $D_n = -\nu k_n^2 -\mu(\delta_{n,0}+\delta_{n,1})$, where $\nu = 2.15 \times 10^{-4}$ is the viscosity and $\mu = 2.10 \times 10^{-5}/\gamma$ is the large-scale drag acting only on the first two modes. We scale $\mu$ with $1/\gamma$ arbitrarily, since we found that the large-scale drag needed to prevent condensation at the last two modes decreased with increasing $\gamma$. We used constant-energy-injection-rate forcing, $f_n = \varepsilon(\delta_{n,n_f} + \delta_{n,n_f+1})/(2 u_n^*) $, with a forcing mode number $n_f = 15$ and energy injection rate $\varepsilon=1$. Runs were initialized with random phases and power-law energy spectra with exponent $-2/3$. Equation \eqref{eq:full_eq} was integrated in time using a fourth order Runge-Kutta method with time step $\Delta t = 5 \times 10^{-4}$. Statistics were collected once steady state was reached.

\begin{figure}
\centering
\includegraphics[width=0.85\textwidth]{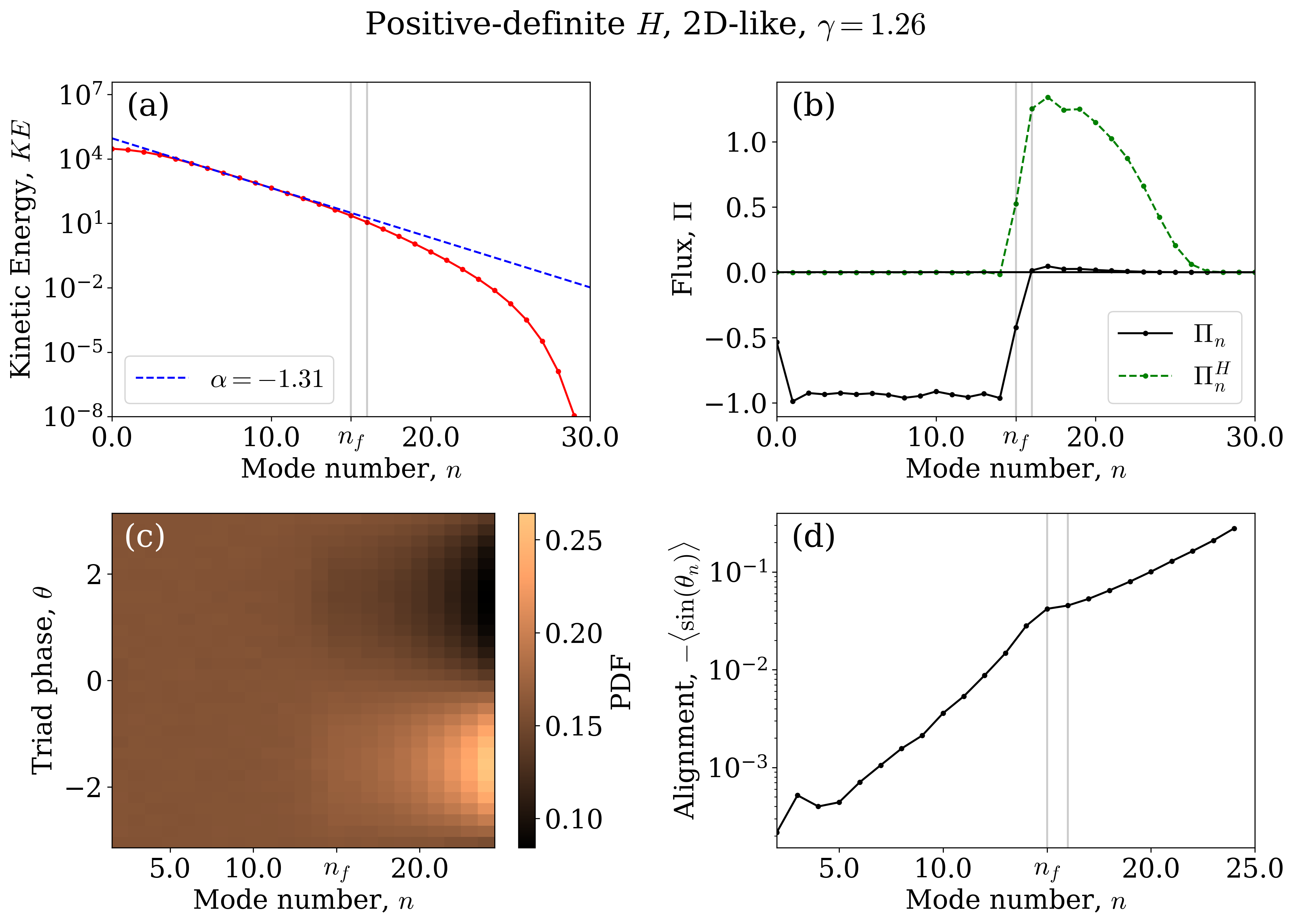}%
\caption{Full 2D-like shell model run for $\gamma=1.26$. In (a) the energy spectrum (averaged at steady state) demonstrates the presence of an inertial range with a power-law energy spectrum containing an exponent close to $\alpha=-1.31$, which we fit with a dashed blue line. In (b) we see that the energy flux is purely inverse and the flux of the second conserved quantity is forward. Panel (c) shows the steady-state PDFs of the triad phases at each mode number, and panel (d) shows the corresponding (negative) alignment associated with the PDFs, revealing a strong scale-dependence. \label{fig:full_run_data}}
\end{figure}
Fig. \ref{fig:full_run_data} shows results from the case of $\gamma = 1.26$, which is representative of all runs for which $0.7 < \gamma < 2$. We find an inverse cascading inertial range for $n<n_f$, with a constant negative flux of energy on average and an energy spectrum with an approximate power-law exponent of $\alpha = -1.31$ (best fit outside of the dissipation ranges and forcing scales). In this inertial range, like that of real 2D fluid systems, there is no flux of enstrophy. The enstrophy has a forward flux to larger $n$ for $n>n_f$, as expected. As seen from the bottom row in Fig. \ref{fig:full_run_data}, the triad phases are aligned such that $\langle \sin(\theta_n)\rangle < 0$ for all $n$. However, the alignment strength varies by two orders of magnitude over the range of $n$ in the simulation (Fig. \ref{fig:full_run_data}(d)). In the range $n<n_f$, the alignment strength seems to increase as a power law of the wavenumber (exponentially in mode number), although understanding why that is is beyond the scope of this work. As noted in the main text, the combination of $n$-dependence of the alignment strength and the power-law energy spectrum results in a constant energy flux, cf. Fig. \ref{fig:full_run_data}(b), given by Eq. \eqref{eq:flux_full}.

\begin{figure}
\centering
\includegraphics[width=0.65\textwidth]{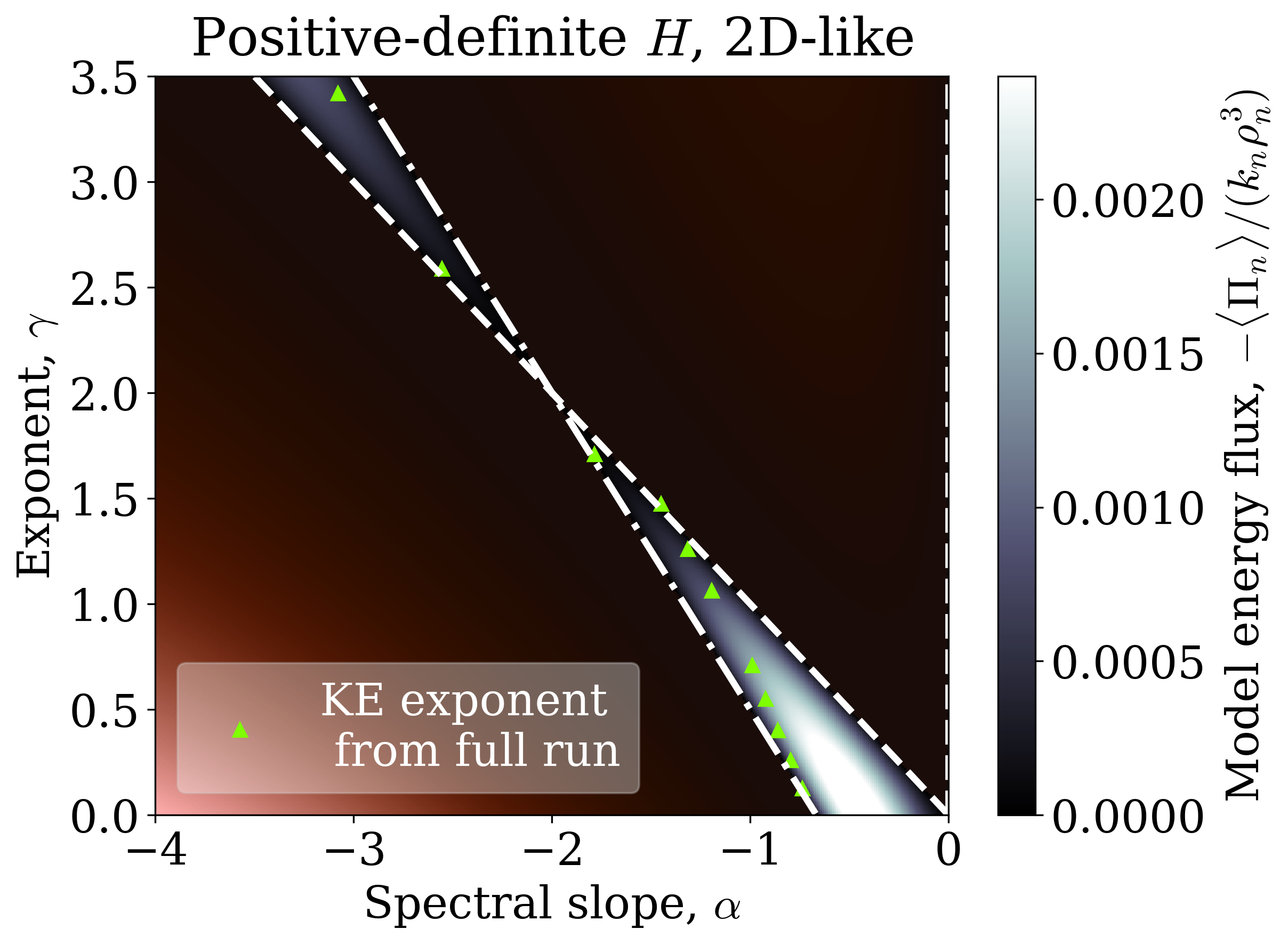}%
\caption{Steady-state energy spectrum exponents of the inverse-cascading inertial range for `full' 2D-like shell model simulations at various values of $\gamma$ plotted over our model's prediction of the rescaled energy flux for a phase-only run with a given value of $(\alpha,\gamma)$. Despite the $n$-independent phase statistics of the phase-only model, which breaks down in the full shell model simulations, the spectrum exponents all lie within the inverse-cascading regions of the parameter space. \label{fig:full_alphas}}
\end{figure}

Between runs, the primary difference observed is the time-averaged kinetic energy power-law exponents for the inverse-cascading inertial range, $\alpha$. Remarkably, the spectral exponents all fall within the regions where inverse energy flux is predicted in our model under the phase-only assumptions (Fig. \ref{fig:full_alphas}). This is rather surprising, given the fact that our zero-flux lines are based on the assumption that the statistics are $n$-independent, yet all of the full runs have clearly $n$-dependent alignment. 

Runs with $\gamma>2$ seem to have somewhat strange behavior in the inertial range for $n<n_f$. It contains two distinct regions: one region of modes (at small $n$, or large scales) seemingly at thermal equilibrium with a flat KE spectrum, and a region with non-zero triad phase alignment and a power-law energy spectrum with the measured slopes found in Fig. \ref{fig:full_alphas}. It seems possible that this is a result of triad phases being influenced by those at $n>n_f$, in the enstrophy cascade, which appears to be operating as expected in these runs. 

Finally, runs with $\gamma < 0.7$ look very similar to those for $0.7 < \gamma < 2$, except for the peculiar fact that there is a non-zero, sustained \textit{inverse} enstrophy cascade for $n<n_f$.

\bibliography{phases_shell}

@article{AlexakisReview,
title = {Cascades and transitions in turbulent flows},
journal = {Physics Reports},
volume = {767-769},
pages = {1-101},
year = {2018},
author = {A. Alexakis and L. Biferale},
}

@article{Cho2001,
author = {Cho, John Y. N. and Lindborg, Erik},
title = {Horizontal velocity structure functions in the upper troposphere and lower stratosphere: 1. Observations},
journal = {Journal of Geophysical Research: Atmospheres},
volume = {106},
number = {D10},
pages = {10223-10232},
year = {2001}
}

@article{Qiu2022,
author = {Qiu, Bo and Nakano, Toshiya and Chen, Shuiming and Klein, Patrice},
title = {Bi-Directional Energy Cascades in the Pacific Ocean From Equator to Subarctic Gyre},
journal = {Geophysical Research Letters},
volume = {49},
number = {8},
pages = {e2022GL097713},
note = {e2022GL097713 2022GL097713},
year = {2022}
}

@article{Balwada2022,
author = {Dhruv Balwada  and Jin-Han Xie  and Raffaele Marino  and Fabio Feraco },
title = {Direct observational evidence of an oceanic dual kinetic energy cascade and its seasonality},
journal = {Science Advances},
volume = {8},
number = {41},
pages = {eabq2566},
year = {2022}}

@article{Sorriso-Valvo2007,
  title = {Observation of Inertial Energy Cascade in Interplanetary Space Plasma},
  author = {Sorriso-Valvo, L. and Marino, R. and Carbone, V. and Noullez, A. and Lepreti, F. and Veltri, P. and Bruno, R. and Bavassano, B. and Pietropaolo, E.},
  journal = {Phys. Rev. Lett.},
  volume = {99},
  issue = {11},
  pages = {115001},
  numpages = {4},
  year = {2007},
  month = {Sep},
  publisher = {American Physical Society},
}

@article{Pestana2019,
  title = {Regime transition in the energy cascade of rotating turbulence},
  author = {Pestana, T. and Hickel, S.},
  journal = {Phys. Rev. E},
  volume = {99},
  issue = {5},
  pages = {053103},
  numpages = {8},
  year = {2019},
  month = {May},
  publisher = {American Physical Society},
  doi = {10.1103/PhysRevE.99.053103},
  url = {https://link.aps.org/doi/10.1103/PhysRevE.99.053103}
}

@article{
Marino2025,
author = {Raffaele Marino  and Jin-Han Xie },
title = {Flux equipartition in astrophysical plasma turbulence},
journal = {Science Advances},
volume = {11},
number = {27},
pages = {eadv8988},
year = {2025}}

@article{Fjortoft1953,
author = {Fj{\o}rtoft, Ragnar},
title = {On the Changes in the Spectral Distribution of Kinetic Energy for Twodimensional, Nondivergent Flow},
journal = {Tellus},
volume = {5},
number = {3},
pages = {225-230},
year = {1953}
}

@article{Kraichnan1967,
    author = {Kraichnan, Robert H.},
    title = {Inertial Ranges in Two‐Dimensional Turbulence},
    journal = {Physics of Fluids},
    volume = {10},
    number = {7},
    pages = {1417-1423},
    year = {1967},
    month = {07},
}

@article{Kraichnan1971, title={Inertial-range transfer in two- and three-dimensional turbulence}, volume={47}, number={3}, journal={Journal of Fluid Mechanics}, author={Kraichnan, Robert H.}, year={1971}, pages={525–535}}

@article{Kraichnan1975, title={Statistical dynamics of two-dimensional flow}, volume={67}, number={1}, journal={Journal of Fluid Mechanics}, author={Kraichnan, Robert H.}, year={1975}, pages={155–175}}

@article{Kraichnan1973, title={Helical turbulence and absolute equilibrium}, volume={59}, number={4}, journal={Journal of Fluid Mechanics}, author={Kraichnan, Robert H.}, year={1973}, pages={745–752}}

@article{Kraichnan1980,
year = {1980},
month = {may},
publisher = {},
volume = {43},
number = {5},
pages = {547},
author = {R H Kraichnan and D Montgomery},
title = {Two-dimensional turbulence},
journal = {Reports on Progress in Physics},
}

@article{Lee1952,
 author = {T. D. Lee},
 journal = {Quarterly of Applied Mathematics},
 number = {1},
 pages = {69--74},
 publisher = {Brown University},
 title = {ON SOME STATISTICAL PROPERTIES OF HYDRODYNAMICAL AND MAGNETO-HYDRODYNAMICAL FIELDS},
 urldate = {2025-03-18},
 volume = {10},
 year = {1952}
}

@article{Hopf1952,
 author = {Eberhard Hopf},
 journal = {Journal of Rational Mechanics and Analysis},
 pages = {87--123},
 publisher = {Indiana University Mathematics Department},
 title = {Statistical Hydromechanics and Functional Calculus},
 urldate = {2025-03-18},
 volume = {1},
 year = {1952}
}

@article{SchimanskyGeier1990,
author={Schimansky-Geier, L.
and Z{\"u}licke, Ch.},
title={Harmonic noise: Effect on bistable systems},
journal={Zeitschrift f{\"u}r Physik B Condensed Matter},
year={1990},
month={Oct},
day={01},
volume={79},
number={3},
pages={451-460},
}

@article{Herbert2014, title={Restricted equilibrium and the energy cascade in rotating and stratified flows}, volume={758}, journal={Journal of Fluid Mechanics}, author={Herbert, Corentin and Pouquet, Annick and Marino, Raffaele}, year={2014}, pages={374–406}}

@article{Moffatt2014, title={Note on the triad interactions of homogeneous turbulence}, volume={741}, journal={Journal of Fluid Mechanics}, author={Moffatt, H. K.}, year={2014}, pages={R3}}

@article{Biferale2017,
    author = {Biferale, L. and Buzzicotti, M. and Linkmann, M.},
    title = {From two-dimensional to three-dimensional turbulence through two-dimensional three-component flows},
    journal = {Physics of Fluids},
    volume = {29},
    number = {11},
    pages = {111101},
    year = {2017},
    month = {07},
}

@article{Biferale2012,
  title = {Inverse Energy Cascade in Three-Dimensional Isotropic Turbulence},
  author = {Biferale, Luca and Musacchio, Stefano and Toschi, Federico},
  journal = {Phys. Rev. Lett.},
  volume = {108},
  issue = {16},
  pages = {164501},
  numpages = {5},
  year = {2012},
  month = {Apr},
  publisher = {American Physical Society},
}

@article{Biferale2013, title={Split energy–helicity cascades in three-dimensional homogeneous and isotropic turbulence}, volume={730}, journal={Journal of Fluid Mechanics}, author={Biferale, L. and Musacchio, S. and Toschi, F.}, year={2013}, pages={309–327}}

@article{Sahoo2017,
  title = {Discontinuous Transition from Direct to Inverse Cascade in Three-Dimensional Turbulence},
  author = {Sahoo, Ganapati and Alexakis, Alexandros and Biferale, Luca},
  journal = {Phys. Rev. Lett.},
  volume = {118},
  issue = {16},
  pages = {164501},
  numpages = {5},
  year = {2017},
  month = {Apr},
  publisher = {American Physical Society},
}

@article{Alexakis2022,
  title = {$\lambda$-Navier–Stokes turbulence},
  author = {Alexakis, Alexandros and Biferale, Luca},
  journal = {Phil. Trans. R. Soc. A.},
  volume = {380},
  pages = {20210243},
  year = {2022},
  publisher = {The Royal Society},
}

@article{VanKan2020, title={Critical transition in fast-rotating turbulence within highly elongated domains}, volume={899}, journal={Journal of Fluid Mechanics}, publisher={Cambridge University Press}, author={van Kan, A. and Alexakis, A.}, year={2020}, pages={A33}}

@article{Benavides2017, title={Critical transitions in thin layer turbulence}, volume={822}, journal={Journal of Fluid Mechanics}, author={Benavides, Santiago Jose and Alexakis, Alexandros}, year={2017}, pages={364–385}}

@article{Seshasayanan2016,
  title = {Critical behavior in the inverse to forward energy transition in two-dimensional magnetohydrodynamic flow},
  author = {Seshasayanan, Kannabiran and Alexakis, Alexandros},
  journal = {Phys. Rev. E},
  volume = {93},
  issue = {1},
  pages = {013104},
  numpages = {13},
  year = {2016},
  month = {Jan},
  publisher = {American Physical Society},
}

@book{DitlevsenBook,
  title={Turbulence and shell models},
  author={Ditlevsen, Peter D},
  year={2010},
  publisher={Cambridge University Press}
}

@article{Vieillefosse1984,
title = {Internal motion of a small element of fluid in an inviscid flow},
journal = {Physica A: Statistical Mechanics and its Applications},
volume = {125},
number = {1},
pages = {150-162},
year = {1984},
issn = {0378-4371},
author = {P. Vieillefosse}
}

@article{Vieillefosse1982,
  title={Local interaction between vorticity and shear in a perfect incompressible fluid},
  author={Vieillefosse, P},
  journal={Journal de Physique},
  volume={43},
  number={6},
  pages={837--842},
  year={1982},
  publisher={Soci{\'e}t{\'e} Fran{\c{c}}aise de Physique}
}

@article{JohnsonReview,
   author = "Johnson, Perry L. and Wilczek, Michael",
   title = "Multiscale Velocity Gradients in Turbulence", 
   journal= "Annual Review of Fluid Mechanics",
   year = "2024",
   volume = "56",
   number = "Volume 56, 2024",
   pages = "463-490",
   publisher = "Annual Reviews",
   issn = "1545-4479",
   type = "Journal Article",
  }

@article{Biferale2003Review,
   author = "Biferale, Luca",
   title = "Shell models of energy cascade in turbulence", 
   journal= "Annual Review of Fluid Mechanics",
   year = "2003",
   volume = "35",
   number = "Volume 35, 2003",
   pages = "441-468",
   publisher = "Annual Reviews",
   issn = "1545-4479",
   type = "Journal Article",
  }

@article{Seshasayanan2014,
  title = {On the edge of an inverse cascade},
  author = {Seshasayanan, Kannabiran and Benavides, Santiago Jose and Alexakis, Alexandros},
  journal = {Phys. Rev. E},
  volume = {90},
  issue = {5},
  pages = {051003},
  numpages = {5},
  year = {2014},
  month = {Nov},
  publisher = {American Physical Society},
}

@article{Alexakis2017, title={Helically decomposed turbulence}, volume={812}, journal={Journal of Fluid Mechanics}, author={Alexakis, Alexandros}, year={2017}, pages={752–770}}

@article{Waleffe1993,
    author = {Waleffe, Fabian},
    title = {Inertial transfers in the helical decomposition},
    journal = {Physics of Fluids A: Fluid Dynamics},
    volume = {5},
    number = {3},
    pages = {677-685},
    year = {1993},
    month = {03},
    issn = {0899-8213},
}

@article{Waleffe1992,
    author = {Waleffe, Fabian},
    title = {The nature of triad interactions in homogeneous turbulence},
    journal = {Physics of Fluids A: Fluid Dynamics},
    volume = {4},
    number = {2},
    pages = {350-363},
    year = {1992},
    month = {02},
    issn = {0899-8213},
}

@article{vanBokhoven2009,
    author = {van Bokhoven, L. J. A. and Clercx, H. J. H. and van Heijst, G. J. F. and Trieling, R. R.},
    title = {Experiments on rapidly rotating turbulent flows},
    journal = {Physics of Fluids},
    volume = {21},
    number = {9},
    pages = {096601},
    year = {2009},
    month = {09},
    issn = {1070-6631},
}

@article{Yarom2013,
    author = {Yarom, Ehud and Vardi, Yuval and Sharon, Eran},
    title = {Experimental quantification of inverse energy cascade in deep rotating turbulence},
    journal = {Physics of Fluids},
    volume = {25},
    number = {8},
    pages = {085105},
    year = {2013},
    month = {08},
    issn = {1070-6631},
}

@article{Smith1999,
    author = {Smith, Leslie M. and Waleffe, Fabian},
    title = {Transfer of energy to two-dimensional large scales in forced, rotating three-dimensional turbulence},
    journal = {Physics of Fluids},
    volume = {11},
    number = {6},
    pages = {1608-1622},
    year = {1999},
    month = {06},
    issn = {1070-6631},
}

@article{Buzzicotti2018,
  title = {Energy transfer in turbulence under rotation},
  author = {Buzzicotti, Michele and Aluie, Hussein and Biferale, Luca and Linkmann, Moritz},
  journal = {Phys. Rev. Fluids},
  volume = {3},
  issue = {3},
  pages = {034802},
  numpages = {22},
  year = {2018},
  month = {Mar},
  publisher = {American Physical Society},
}

@article{Benzi1993,
title = {On intermittency in a cascade model for turbulence},
journal = {Physica D: Nonlinear Phenomena},
volume = {65},
number = {1},
pages = {163-171},
year = {1993},
issn = {0167-2789},
author = {R. Benzi and L. Biferale and G. Parisi}
}

@article{Buzzicotti2016,
author={Buzzicotti, Michele
and Murray, Brendan P.
and Biferale, Luca
and Bustamante, Miguel D.},
title={Phase and precession evolution in the Burgers equation},
journal={The European Physical Journal E},
year={2016},
month={Mar},
day={25},
volume={39},
number={3},
pages={34},
}

@article{Murray2018, title={Energy flux enhancement, intermittency and turbulence via Fourier triad phase dynamics in the 1-D Burgers equation}, volume={850},journal={Journal of Fluid Mechanics}, author={Murray, Brendan P. and Bustamante, Miguel D.}, year={2018}, pages={624–645}}

@article{Protas2024,
  title = {Alignments of triad phases in extreme one-dimensional Burgers flows},
  author = {Protas, Bartosz and Kang, Di and Bustamante, Miguel D.},
  journal = {Phys. Rev. E},
  volume = {109},
  issue = {5},
  pages = {055104},
  numpages = {15},
  year = {2024},
  month = {May},
  publisher = {American Physical Society},
}

@article{Arguedas2022,
  title = {Minimal phase-coupling model for intermittency in turbulent systems},
  author = {Arguedas-Leiva, Jos\'e-Agust\'{\i}n and Carroll, Enda and Biferale, Luca and Wilczek, Michael and Bustamante, Miguel D.},
  journal = {Phys. Rev. Res.},
  volume = {4},
  issue = {3},
  pages = {L032035},
  numpages = {6},
  year = {2022},
  month = {Aug},
  publisher = {American Physical Society},
}

@article{Wang2024,
  title = {Role of Fourier phase dynamics in decaying turbulence},
  author = {Wang, Chuhan and Fang, Le and Wang, Zhan and Xu, Chunxiao},
  journal = {Phys. Rev. Fluids},
  volume = {9},
  issue = {11},
  pages = {114603},
  numpages = {12},
  year = {2024},
  month = {Nov},
  publisher = {American Physical Society},
}

@misc{Kang2021,
      title={Alignments of Triad Phases in 1D Burgers and 3D Navier-Stokes Flows}, 
      author={Di Kang and Bartosz Protas and Miguel D. Bustamante},
      year={2021},
      eprint={2105.09425},
      archivePrefix={arXiv},
      primaryClass={physics.flu-dyn},
}

@article{Reynolds2016,
year = {2016},
month = {sep},
publisher = {EDP Sciences, IOP Publishing and Società Italiana di Fisica},
volume = {115},
number = {3},
pages = {34002},
author = {Reynolds-Barredo, J. M. and Newman, D. E. and Terry, P. W. and Sanchez, R.},
title = {Fourier signature of filamentary vorticity structures in two-dimensional turbulence},
journal = {Europhysics Letters}
}

@article{MarstonReview,
   author = "Marston, J.B. and Tobias, S.M.",
   title = "Recent Developments in Theories of Inhomogeneous and Anisotropic Turbulence", 
   journal= "Annual Review of Fluid Mechanics",
   year = "2023",
   volume = "55",
   number = "Volume 55, 2023",
   pages = "351-375",
   publisher = "Annual Reviews",
   issn = "1545-4479",
   type = "Journal Article",
  }

@article{Ballouz2020,
  title = {Geometric constraints on energy transfer in the turbulent cascade},
  author = {Ballouz, Joseph G. and Ouellette, Nicholas T.},
  journal = {Phys. Rev. Fluids},
  volume = {5},
  issue = {3},
  pages = {034603},
  numpages = {14},
  year = {2020},
  month = {Mar},
  publisher = {American Physical Society},
}

@article{Chen2006,
  title = {Physical Mechanism of the Two-Dimensional Inverse Energy Cascade},
  author = {Chen, Shiyi and Ecke, Robert E. and Eyink, Gregory L. and Rivera, Michael and Wan, Minping and Xiao, Zuoli},
  journal = {Phys. Rev. Lett.},
  volume = {96},
  issue = {8},
  pages = {084502},
  numpages = {4},
  year = {2006},
  month = {Feb},
  publisher = {American Physical Society}
}

@article{Storer2022,
author={Storer, Benjamin A.
and Buzzicotti, Michele
and Khatri, Hemant
and Griffies, Stephen M.
and Aluie, Hussein},
title={Global energy spectrum of the general oceanic circulation},
journal={Nature Communications},
year={2022},
month={Sep},
day={09},
volume={13},
number={1},
pages={5314},
issn={2041-1723},
}

@book{VermaBook,
  title={Energy transfers in fluid flows: multiscale and spectral perspectives},
  author={Verma, Mahendra K},
  year={2019},
  publisher={Cambridge University Press}
}

@article{Mcwilliams1990, title={The vortices of two-dimensional turbulence}, volume={219}, journal={Journal of Fluid Mechanics}, author={Mcwilliams, James C.}, year={1990}, pages={361–385}}

@article{Eyink20062DCL, title={A turbulent constitutive law for the two-dimensional inverse energy cascade}, volume={549}, journal={Journal of Fluid Mechanics}, author={Eyink, Gregory L.}, year={2006}, pages={191–214}}

@article{Xiao2009, title={Physical mechanism of the inverse energy cascade of two-dimensional turbulence: a numerical investigation}, volume={619}, journal={Journal of Fluid Mechanics}, author={Xiao, Z. and Wan, M. and Chen, S. and Eyink, G. L.}, year={2009}, pages={1–44}}

@article{Ballouz2018, title={Tensor geometry in the turbulent cascade}, volume={835}, journal={Journal of Fluid Mechanics}, author={Ballouz, Joseph G. and Ouellette, Nicholas T.}, year={2018}, pages={1048–1064}}

@article{Park2025,
author={Park, Danah
and Lozano-Dur{\'a}n, Adri{\'a}n},
title={The coherent structure of the energy cascade in isotropic turbulence},
journal={Scientific Reports},
year={2025},
month={Jan},
day={02},
volume={15},
number={1},
pages={14},
issn={2045-2322},
}

@article{Eyink2006MSG, title={Multi-scale gradient expansion of the turbulent stress tensor}, volume={549}, journal={Journal of Fluid Mechanics}, author={Eyink, Gregory L.}, year={2006}, pages={159–190}}

@article{Doan2018,
  title = {Scale locality of the energy cascade using real space quantities},
  author = {Doan, N. A. K. and Swaminathan, N. and Davidson, P. A. and Tanahashi, M.},
  journal = {Phys. Rev. Fluids},
  volume = {3},
  issue = {8},
  pages = {084601},
  numpages = {14},
  year = {2018},
  month = {Aug},
  publisher = {American Physical Society}
}

@article{Johnson2020,
  title = {Energy Transfer from Large to Small Scales in Turbulence by Multiscale Nonlinear Strain and Vorticity Interactions},
  author = {Johnson, Perry L.},
  journal = {Phys. Rev. Lett.},
  volume = {124},
  issue = {10},
  pages = {104501},
  numpages = {6},
  year = {2020},
  month = {Mar},
  publisher = {American Physical Society}
}

@article{Johnson2021, title={On the role of vorticity stretching and strain self-amplification in the turbulence energy cascade}, volume={922}, journal={Journal of Fluid Mechanics}, author={Johnson, Perry L.}, year={2021}, pages={A3}}

@article{Carbone2020, title={Is vortex stretching the main cause of the turbulent energy cascade?}, volume={883}, journal={Journal of Fluid Mechanics}, author={Carbone, M. and Bragg, A. D.}, year={2020}, pages={R2}}

@article{Jimenez2021,
  title = {Collective organization and screening in two-dimensional turbulence},
  author = {Jim\'enez, Javier},
  journal = {Phys. Rev. Fluids},
  volume = {6},
  issue = {8},
  pages = {084601},
  numpages = {24},
  year = {2021},
  month = {Aug},
  publisher = {American Physical Society},
}

@article{Wunsch2004,
   author = "Wunsch, Carl and Ferrari, Raffaele",
   title = "VERTICAL MIXING, ENERGY, AND THE GENERAL CIRCULATION OF THE OCEANS", 
   journal= "Annual Review of Fluid Mechanics",
   year = "2004",
   volume = "36",
   number = "Volume 36, 2004",
   pages = "281-314",
   publisher = "Annual Reviews",
   issn = "1545-4479",
   type = "Journal Article",
  }

@article{Mellor1982,
author = {Mellor, George L. and Yamada, Tetsuji},
title = {Development of a turbulence closure model for geophysical fluid problems},
journal = {Reviews of Geophysics},
volume = {20},
number = {4},
pages = {851-875},
year = {1982}
}

@book{KalnayBook,
  title={Atmospheric modeling, data assimilation and predictability},
  author={Kalnay, Eugenia},
  year={2003},
  publisher={Cambridge university press}
}

@article{Dauxois2021,
  title = {Confronting Grand Challenges in environmental fluid mechanics},
  author = {Dauxois, T. and Peacock, T. and Bauer, P. and Caulfield, C. P. and Cenedese, C. and Gorl\'e, C. and Haller, G. and Ivey, G. N. and Linden, P. F. and Meiburg, E. and Pinardi, N. and Vriend, N. M. and Woods, A. W.},
  journal = {Phys. Rev. Fluids},
  volume = {6},
  issue = {2},
  pages = {020501},
  numpages = {40},
  year = {2021},
  month = {Feb},
  publisher = {American Physical Society},
}

@article{vanKan2024,
    author = {van Kan, Adrian},
    title = {Phase transitions in anisotropic turbulence},
    journal = {Chaos: An Interdisciplinary Journal of Nonlinear Science},
    volume = {34},
    number = {12},
    pages = {122103},
    year = {2024},
    month = {12},
    issn = {1054-1500},
}

@article{BoffettaReview,
   author = "Boffetta, Guido and Ecke, Robert E.",
   title = "Two-Dimensional Turbulence", 
   journal= "Annual Review of Fluid Mechanics",
   year = "2012",
   volume = "44",
   number = "Volume 44, 2012",
   pages = "427-451",
   publisher = "Annual Reviews",
   issn = "1545-4479",
   type = "Journal Article",
  }

@article{JimenezReview,
   author = "Jiménez, Javier",
   title = "Cascades in Wall-Bounded Turbulence", 
   journal= "Annual Review of Fluid Mechanics",
   year = "2012",
   volume = "44",
   number = "Volume 44, 2012",
   pages = "27-45",
   publisher = "Annual Reviews",
   issn = "1545-4479",
   type = "Journal Article",
  }

@book{DavidsonBook,
	title={Turbulence in rotating, stratified and electrically conducting fluids},
	author={Davidson, P. A.},
	year={2013},
	publisher={Cambridge University Press}
}

@book{FrischBook,
	title={Turbulence: the legacy of AN Kolmogorov},
	author={Frisch, U.},
	year={1995},
	publisher={Cambridge University Press}
}

@article{Moradi2017,
    author = {Moradi, Sara and Teaca, Bogdan and Anderson, Johan},
    title = {Role of phase synchronisation in turbulence},
    journal = {AIP Advances},
    volume = {7},
    number = {11},
    pages = {115213},
    year = {2017},
    month = {11},
}

@article{Bourouiba2008,
    author = {Bourouiba, L.},
    title = {Model of a truncated fast rotating flow at infinite Reynolds number},
    journal = {Physics of Fluids},
    volume = {20},
    number = {7},
    pages = {075112},
    year = {2008},
    month = {07},
}

@article{Lvov1998,
  title = {Improved shell model of turbulence},
  author = {L'vov, Victor S. and Podivilov, Evgenii and Pomyalov, Anna and Procaccia, Itamar and Vandembroucq, Damien},
  journal = {Phys. Rev. E},
  volume = {58},
  issue = {2},
  pages = {1811--1822},
  numpages = {0},
  year = {1998},
  month = {Aug},
  publisher = {American Physical Society}
}

@article{Aurell1994,
  title = {Statistical mechanics of shell models for two-dimensional turbulence},
  author = {Aurell, E. and Boffetta, G. and Crisanti, A. and Frick, P. and Paladin, G. and Vulpiani, A.},
  journal = {Phys. Rev. E},
  volume = {50},
  issue = {6},
  pages = {4705--4715},
  numpages = {0},
  year = {1994},
  month = {Dec},
  publisher = {American Physical Society}
}

@article{Gurcan2016,
  title = {Logarithmic discretization and systematic derivation of shell models in two-dimensional turbulence},
  author = {G\"urcan, \"O. D. and Morel, P. and Kobayashi, S. and Singh, Rameswar and Xu, S. and Diamond, P. H.},
  journal = {Phys. Rev. E},
  volume = {94},
  issue = {3},
  pages = {033106},
  numpages = {10},
  year = {2016},
  month = {Sep},
  publisher = {American Physical Society},
}

@book{CarrollThesis,
  title={Fourier phase synchronization in minimal models for turbulence},
  author={Carroll, E.},
  series={Ph.D. Thesis},
  year={2023},
  volume={15301},
  publisher={University College Dublin}
}

@software{Benavides2025Code,
  author       = {Santiago J. Benavides},
  title        = {s-benavides/MHD\_shell\_model: MHD\_shell\_model v1.0Z
                   (Z for zenodo)
                  },
  month        = jul,
  year         = 2025,
  publisher    = {Zenodo},
  note         = {\url{https://doi.org/10.5281/zenodo.15800547}},
  version      = {v1.0Z},
  doi          = {10.5281/zenodo.15800547},
  url          = {https://doi.org/10.5281/zenodo.15800547},
  swhid        = {swh:1:dir:484e56a7b09505f0546b43db574bf8b5f4b18c33
                   ;origin=https://doi.org/10.5281/zenodo.15800546;vi
                   sit=swh:1:snp:8956b77482357628f2b2f502cd329636c3f7
                   fc7f;anchor=swh:1:rel:2378a4a3487e6b852200a065ae14
                   049150e0eb30;path=s-benavides-
                   MHD\_shell\_model-208a9ec
                  },
}

@software{Benavides2025Data,
author = {Santiago J. Benavides and Miguel D. Bustamante},
title = {Data and scripts for figures in {B}enavides \& {B}ustamante ``{P}hase dynamics and their role determining energy flux in hydrodynamic shell models.'' (2025).},
year = {2025},
month = {7},
url = {https://doi.org/10.6084/m9.figshare.29469659},
note = {\url{https://doi.org/10.6084/m9.figshare.29469659}},
doi = {10.6084/m9.figshare.29469659},
}

@misc{Manfredini2025,
      title={Nonlinear phase synchronization and the role of spacing in shell models}, 
      author={Lorenzo Manfredini and Özgür D. Gürcan},
      year={2025},
      eprint={2507.14142},
      archivePrefix={arXiv},
      primaryClass={nlin.CD},
}

@book{VanKampen1992,
	title={Stochastic processes in physics and chemistry},
	author={Van Kampen, Nicolaas Godfried},
	volume={1},
	year={1992},
	publisher={Elsevier}
}

@misc{Weisstein,
    author   = {Weisstein, Eric W.},
    title    = {Modified Bessel Function of the First Kind. {From MathWorld---A Wolfram Web Resource}},
    url      = {https://mathworld.wolfram.com/ModifiedBesselFunctionoftheFirstKind.html},
    note     = {\url{https://mathworld.wolfram.com/ModifiedBesselFunctionoftheFirstKind.html}. Last visited on 3/7/2025.}
}

@article{Mailybaev2021,
  title = {Hidden scale invariance of intermittent turbulence in a shell model},
  author = {Mailybaev, Alexei A.},
  journal = {Phys. Rev. Fluids},
  volume = {6},
  issue = {1},
  pages = {L012601},
  numpages = {7},
  year = {2021},
  month = {Jan},
  publisher = {American Physical Society},
}

@article{Mailybaev2022,
  title = {Shell model intermittency is the hidden self-similarity},
  author = {Mailybaev, Alexei A.},
  journal = {Phys. Rev. Fluids},
  volume = {7},
  issue = {3},
  pages = {034604},
  numpages = {19},
  year = {2022},
  month = {Mar},
  publisher = {American Physical Society},
}

@article{deWit2024,
  title = {Extreme statistics and extreme events in dynamical models of turbulence},
  author = {de Wit, Xander M. and Ortali, Giulio and Corbetta, Alessandro and Mailybaev, Alexei A. and Biferale, Luca and Toschi, Federico},
  journal = {Phys. Rev. E},
  volume = {109},
  issue = {5},
  pages = {055106},
  numpages = {9},
  year = {2024},
  month = {May},
  publisher = {American Physical Society},
}

@article{Ditlevsen1996,
  title = {Cascades and statistical equilibrium in shell models of turbulence},
  author = {Ditlevsen, P. D. and Mogensen, I. A.},
  journal = {Phys. Rev. E},
  volume = {53},
  issue = {5},
  pages = {4785--4793},
  numpages = {0},
  year = {1996},
  month = {May},
  publisher = {American Physical Society}
}

@article{Gilbert2002,
  title = {Inverse Cascade Regime in Shell Models of Two-Dimensional Turbulence},
  author = {Gilbert, Thomas and L'vov, Victor S. and Pomyalov, Anna and Procaccia, Itamar},
  journal = {Phys. Rev. Lett.},
  volume = {89},
  issue = {7},
  pages = {074501},
  numpages = {4},
  year = {2002},
  month = {Jul},
  publisher = {American Physical Society},
}

@Article{Eyink2003,
author={Eyink, Gregory L.
and Chen, Shiyi
and Chen, Qiaoning},
title={Gibbsian Hypothesis in Turbulence},
journal={Journal of Statistical Physics},
year={2003},
month={Dec},
day={01},
volume={113},
number={5},
pages={719-740},
issn={1572-9613},
}

\end{document}